\documentclass[final]{IEEEtran}
\usepackage{cite}
\usepackage{graphicx}
\usepackage{picinpar}
\usepackage[cmex10]{amsmath}
\usepackage{amsmath,amsfonts,amssymb}
\usepackage[justification=centering]{caption}

\usepackage{algorithmic}
\usepackage{subfigure}
\usepackage[ruled,lined,boxed]{algorithm2e}

\usepackage{stfloats}
\usepackage{bm}
\usepackage{color}
\usepackage{stfloats}
\usepackage{url}
\usepackage{caption} 
\usepackage{makecell}
\usepackage{multirow}
\usepackage{array}
\usepackage{booktabs}
\usepackage{geometry}
\begin{document}
	\pdfoptionpdfminorversion=6
	\newtheorem{lemma}{Lemma}
	\newtheorem{corol}{Corollary}
	\newtheorem{theorem}{Theorem}
	\newtheorem{proposition}{Proposition}
	\newtheorem{definition}{Definition}
	\newcommand{\e}{\begin{equation}}
		\newcommand{\ee}{\end{equation}}
	\newcommand{\eqn}{\begin{eqnarray}}
		\newcommand{\eeqn}{\end{eqnarray}}
	\renewcommand{\algorithmicrequire}{ \textbf{Input:}} 
	\renewcommand{\algorithmicensure}{ \textbf{Output:}} 
	
\title{Deep Joint Semantic Coding and Beamforming for Near-Space Airship-Borne Massive MIMO Network}
	
\author{Minghui Wu, Zhen Gao, Zhaocheng Wang, Dusit Niyato, George~K.~Karagiannidis, and Sheng Chen
\thanks{M.~Wu and Z.~Gao are with School of Information and Electronics, Beijing Institute of Technology, Beijing 100081, China (e-mails: wuminghui@bit.edu.cn; gaozhen16@bit.edu.cn).} %
\thanks{Z. Wang is with Beijing National Research Center for Information Science and Technology, Department of Electronic Engineering, Tsinghua University, Beijing 100084, China, and Z. Wang is also with Tsinghua Shenzhen International Graduate School, Shenzhen 518055, China (e-mail: zcwang@tsinghua.edu.cn).}

\thanks{Dusit Niyato is with the School of Computer Science and Engineering, Nanyang Technological University, Singapore 639798, Singapore (e-mail: dniyato@ntu.edu.sg).}
\thanks{G. K. Karagiannidis is with with the Department of Electrical and Computer Engineering, Aristotle University of Thessaloniki, 54124 Thessaloniki, Greece and also with the Artificial Intelligence \& Cyber Systems Research Center, Lebanese American University (LAU), Lebanon (e-mail: geokarag@auth.gr).}
\thanks{Sheng Chen is with the School of Electronics and Computer Science, University of Southampton, Southampton SO17 1BJ, U.K. (e-mail: sqc@soton.ac.uk).} %
\vspace*{-5mm}
}
\maketitle
	
\begin{abstract}
Near-space airship-borne communication network is recognized to be an indispensable component of the future  integrated ground-air-space network thanks to airships' advantage of long-term residency at stratospheric altitudes, but it urgently needs reliable and efficient Airship-to-X link. To improve the transmission efficiency and capacity, this paper proposes to integrate semantic communication with massive multiple-input multiple-output (MIMO) technology. 
Specifically, we propose a deep joint semantic coding and beamforming (JSCBF) scheme for airship-based massive MIMO image transmission network in space, in which semantics from both source and channel are fused to jointly design the semantic coding and physical layer beamforming. First, we design two semantic extraction networks to extract semantics from image source and channel state information, respectively. Then, we propose a semantic fusion network that can fuse these semantics into complex-valued semantic features for subsequent physical-layer transmission. To efficiently transmit the fused semantic features at the physical layer, we then propose the hybrid data and model-driven semantic-aware beamforming networks. At the receiver, a semantic decoding network is designed to reconstruct the transmitted images. Finally, we perform end-to-end deep learning to jointly train all the modules, using the image reconstruction quality at the receivers as a metric. The proposed deep JSCBF scheme fully combines the efficient source compressibility and robust error correction capability of semantic communication with the high spectral efficiency of massive MIMO, achieving a significant performance improvement over existing approaches.
\end{abstract}
	
\begin{IEEEkeywords}
Airship base station, beamforming, massive MIMO, deep learning, semantic communication.
\end{IEEEkeywords}

\IEEEpeerreviewmaketitle
	
\section{Introduction}\label{S1}

\begin{figure}[t]
	\centering
	\includegraphics[width = 1 \columnwidth,keepaspectratio]{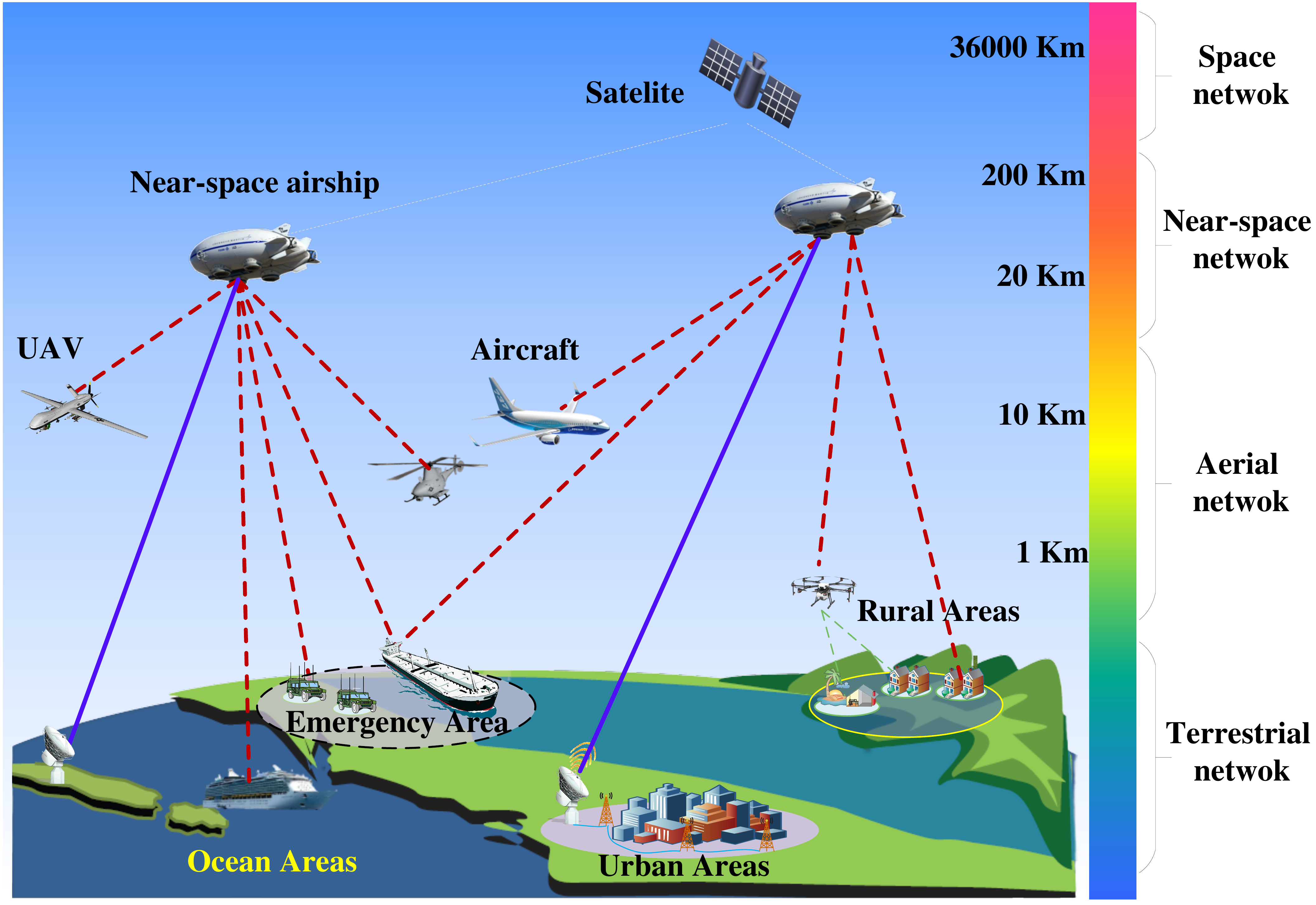}
	\captionsetup{font={footnotesize}, singlelinecheck = off, justification = raggedright,name={Fig.},labelsep=period}
	\caption{Schematic diagram of the IGASN architecture, consisting of space satellites, near-space airships, aerial aircraft, and terrestrial BSs.}
	\label{fig:SYS} 
	\vspace*{-4mm}
\end{figure}

In the evolving 6G communications landscape, the integrated ground-air-space network (IGASN), as shown in Fig.~\ref{fig:SYS}, is increasingly recognized as a key architecture. Within this framework, the incorporation of an airship-based near-space communications network is seen as a breakthrough extension, offering revolutionary communications capabilities and extending the coverage and efficiency of the network \cite{IGASN-1}. Unlike space-based satellite communication systems, which offer wide coverage but suffer from long transmission latency and high costs due to their higher orbital positions, airship-based near-space base stations (BSs) can remain continuously in the stratosphere, closer to Earth, for several years and can be flexibly deployed as needed. This feature provides reduced transmission delays, relatively stable channel propagation conditions, and stronger task-oriented capability for the airship-borne near-space network. Unlike airborne platforms such as unmanned aerial vehicles, which are constrained by limited coverage, shorter endurance and regulatory restrictions, airship-based BSs provide broader and uninterrupted network services. In addition, unlike terrestrial BSs that may have obstructed coverage due to geographic features or infrastructure limitations, airship-based BSs can provide unobstructed wide-area coverage \cite{JSAC_Wu1}. This attribute is particularly beneficial in areas where terrestrial networks are either non-existent, inefficient, damaged or overwhelmed by demand. Given that approximately 70\% of the Earth's surface is covered by water, these airship-based BSs are uniquely positioned to provide coverage in vast maritime regions and remote areas beyond the reach of terrestrial BSs. They also offer enhanced robustness and resilience in regions covered by terrestrial BSs, serving as a critical backup for communications support, particularly when ground infrastructure is compromised by natural disasters or other emergencies.

However, the deployment of airship-based near-space communication networks faces the challenge of massive data transmission demands with limited transmission resources. As a novel paradigm in 6G networks, semantic communication can effectively mitigate this problem by achieving higher transmission efficiency \cite{Sem_intro1, Sem_intro2, Sem_intro3}. Building on Weaver and Shannon's definition of semantic information \cite{Sem_intro4}, this innovative paradigm shifts the focus to the underlying meaning of symbols rather than the pursuit of precise reconstruction. Semantic communication systems, unlike their traditional counterparts, can significantly compress source information and reduce associated communication costs. In scenarios where conveying intrinsic meaning is the primary objective, semantic communication will play an indispensable role.

Massive multiple-input multiple-output (MIMO) is a key technology for future wireless communication systems \cite{MIMO1}. By employing large antenna arrays on the airborne BSs coupled with advanced beamforming techniques, the transmission capacity can be significantly increased to meet the massive data transmission requirements of the IGASN. Therefore, the integration of massive MIMO beamforming and semantic communication techniques in an airship-based near-space communication network represents a very promising communication paradigm by simultaneously exploiting the ability of semantic communication to extract key semantics from the source and the ability of massive MIMO to handle high data rate transmissions for 6G massive communication.

\subsection{State-of the-Art}\label{S1.1}

In the field of semantic communication, the success of deep learning (DL) has inspired the adoption of architectures based on deep neural networks (DNN), such as autoencoders, which have been widely used in semantic communication systems to achieve better performance \cite{Sem_intro3}. Current research in semantic communication is mainly divided into two different directions: the first one focuses on the design of efficient semantic encoding and decoding algorithms \cite{Deep-JSCC,NTSCC1,NTSCC2,Speech1,Speech2,R1-6}, while the second one is dedicated to the development of algorithms for the physical layer transmission modules of semantic communication \cite{R1-14,R1-15,R1-19,R1-16,R1-17,multiModal-sem,Sem_MIMO,Vit_MIMO}.

In the research of semantic coding and decoding algorithms, the transmitter performing semantic coding and the receiver performing semantic decoding are considered as a pair of DNN-based autoencoders, where the encoder at the transmitter semantically extracts and encodes the source data into a complex-valued transmit signal, and the decoder at the receiver decodes the data based on the received symbols. 
For example, deep joint source-channel coding (JSCC) and its improved versions have been proposed in \cite{Deep-JSCC,NTSCC1,NTSCC2}. In these approaches, images are mapped into complex-valued transmission symbols by a semantic encoder and reconstructed by a semantic decoder to achieve improved reconstruction performance. The authors of \cite{Speech1,Speech2} designed semantic communication systems based on the attention mechanism for speech transmission, while the authors of \cite{R1-6} proposed the deep JSCC algorithm for multimodal data transmission.

Compared to the first category of research, which focuses on the design of semantic encoding and decoding without considering actual communication scenarios, the second category pays more attention to the enhancement of physical layer transmission modules in various semantic communication scenarios \cite{R1-14}.
The authors of \cite{R1-15,R1-19} studied resource allocation in semantic communication systems. The work \cite{R1-16} investigated the integration of Orthogonal Frequency Division Multiplexing (OFDM) with deep JSCC and optimized the entire semantic communication system in an end-to-end (E2E) manner. The paper \cite{R1-17} designed a semantic-driven constellation to improve the image reconstruction quality in deep JSCC. 
The authors of \cite{multiModal-sem,Sem_MIMO,Vit_MIMO} further explored the design of physical layer transmission modules in MIMO semantic communication systems. Specifically, the \cite{multiModal-sem} research focuses on semantic transmission in multi-user MIMO uplink systems, while the \cite{Sem_MIMO,Vit_MIMO} research focuses on optimizing power allocation and adaptive channel state information (CSI) feedback code length, respectively, for single-user narrowband MIMO semantic communication systems in the downlink. However, there is currently a lack of research focused on optimizing the downlink multi-user beamforming modules in massive MIMO semantic communication systems.

Traditional non-iterative MIMO beamforming schemes, such as regularized zero forcing (RZF) \cite{R2-4} and signal-to-leakage-and-noise ratio (SLNR) \cite{R2-5} algorithms, are easy to implement, but they tend to achieve suboptimal performance because they do not directly maximize spectral efficiency. To maximize spectral efficiency, the authors of \cite{R1-7,R1-8,R1-9} designed efficient iterative beamforming algorithms based on semidefinite relaxation \cite{R1-7}, weighted minimum mean square error (WMMSE) \cite{R1-8}, and penalty dual decomposition \cite{R1-9} optimization frameworks. Although these iterative beamforming algorithms offer performance close to theoretical limits, their high computational complexity due to large matrix inversions and high number of iterations, coupled with the dependence on accurate CSI information, hinders their application in multi-user massive MIMO systems under imperfect CSI and unfavorable channel propagation conditions.

Motivated by the success of data-driven DL in various fields, its application to massive MIMO systems has also been actively explored in recent years \cite{R2-8}. The authors of \cite{R1-12,R1-13,R2-13} used convolutional neural networks (CNNs) with supervised training to approximate traditional iterative beamforming algorithms. The study \cite{R2-12} proposed a beamforming neural network training strategy based on transfer learning for different channel conditions. The authors of \cite{JSAC_Wu1,My1-46,My1-48,E2E-Pre2} further considered modeling the channel acquisition process and beamforming as an E2E neural network that is jointly trained with the spectral efficiency metric to achieve high spectral efficiency beamforming with limited pilot overhead. These data-driven DL beamforming approaches treat the communication process as a black box and optimize the mapping between inputs and outputs through E2E DL training. This methodology has the advantage of not relying on existing expert knowledge, as it learns transmission strategies directly from data samples, providing better adaptation to imperfect CSI. 
However, these approaches have limited interpretability and generalization capabilities, and their performance may not be guaranteed.

Consequently, model-driven DL-MIMO beamforming techniques that incorporate expert knowledge have attracted considerable attention in recent years.
By unfolding the WMMSE iterative process into a hierarchical network, the authors of \cite{DL_unfloding1,DL_unfloding2} developed an efficient iterative design with neural networks for the narrowband MIMO scenario. By integrating traditional successive over-relaxation-based beamforming schemes with DL, the work \cite{shicong} accelerated the network convergence speed and reduced the number of iterations required. The study \cite{JSAC_Wu2} derived a WMMSE algorithm in a rate-splitting multiple access scenario and combined it with DNNs to achieve better performance. However, model-driven DL relies on accurate expert knowledge and may still suffer performance degradation if the expert knowledge does not match the actual channel conditions.

\subsection{ Motivation and Contribution}\label{S1.2}

While existing semantic communication schemes have been explored for MIMO systems, they rely on traditional downlink beamforming techniques such as singular value decomposition \cite{Sem_MIMO,Vit_MIMO}, which requires accurate CSI. In the context of massive MIMO systems, the constraints of limited pilot overhead inevitably lead to inaccurate CSI estimation, posing significant challenges to achieving good performance. In addition, current approaches are mainly designed for single-user narrowband MIMO systems, leaving a research gap in the application of semantic communication in the more complex multi-user massive MIMO systems. Therefore, further research is needed to develop efficient semantic communication schemes tailored for the multi-user case in near-space airship-based broadband massive MIMO communication systems. This motivates our work.

The discussion of DL-based MIMO beamforming schemes in the previous subsection shows that both data-driven and model-driven DL strategies can effectively improve beamforming performance. Each has its own advantages and limitations, making them suitable for different scenarios. This motivates us to merge the strengths of both approaches by introducing a hybrid data-driven and model-driven beamforming strategy for airship-based multi-user massive MIMO systems, with the aim of achieving superior performance. Furthermore, we integrate this beamforming strategy with semantic communication, specifically focusing on the task of image compression and reconstruction. This integration exploits the semantics of both source and channel to design the semantic coding and physical layer beamforming, culminating in a deep Joint Semantic Coding and Beamforming (JSCBF) scheme for airship-based massive MIMO image transmission network in near space.
  

This paper proposes a deep JSCBF approach for an airship-based massive MIMO image transmission network in near space.
Our research focuses on the image modality, which was chosen for its distinctive ability to showcase the efficiency and potential of semantic transmission techniques. The inherent richness of semantic information and the superior compressibility of the image modality provide an exemplary platform to illustrate the effectiveness of semantic transmission. This research lays the foundation for future extensions to more complex modalities, such as video. The main contributions of this work are summarized below.

\begin{itemize}
	\item To the best of our knowledge, the proposed deep JSCBF scheme is the first to integrate semantic communication with downlink multi-user broadband massive MIMO beamforming. In this scheme, the semantics derived from both images and CSI are jointly used to guide the design of semantic coding and beamforming.
	
	\item For semantic coding, we adopt the transformer architecture to extract semantic features from both images and CSI. In addition, we design a semantic fusion network to fuse the CSI semantics and image semantics to obtain the fused semantic features for the subsequent physical layer transmission. By fusing CSI semantics and image semantics, our deep JSCBF scheme can make the fused semantic features to better adapt to different channel conditions and mitigate the damage to semantic information caused by uncertainties in physical layer transmission.
	
	\item To better transmit the fused semantic features in the physical layer link, we propose a hybrid data-driven and model-driven semantic-aware beamforming scheme. The data-driven semantic-aware beamforming network takes the fused semantic features as input to effectively exploit the embedded image and CSI semantics, and ultimately outputs transmission signals in the dimensionality of the BS antennas. To further ensure the performance of the network and improve its interpretability, we first derive a new beamforming algorithm based on WMMSE for multi-user massive MIMO systems under imperfect CSI. Then, we integrate this derived WMMSE beamforming algorithm with DL to propose a model-driven semantic-aware beamforming network. In this network, a transformer is used to replace the iterative parameter update process in WMMSE. Finally, by weighting and summing the outputs of both beamforming networks, we obtain the final transmission signal, effectively combining the advantages of both data-driven and model-driven DL for improved performance.
	
	\item By adopting a combined loss function that integrates pixel-level distortion loss, i.e., mean square error (MSE), with perceptual metrics, i.e., multi-scale structural similarity (MS-SSIM) and learned perceptual image patch similarity (LPIPS), we achieve the E2E joint training of the proposed networks. Therefore, our deep JSCBF scheme achieves the joint optimization of semantic encoding/decoding and massive MIMO beamforming, resulting in a significant improvement in image reconstruction performance.
\end{itemize}

\textit{Notation}: We use lower-case letters for scalars, lower-case boldface letters for column vectors, and upper-case boldface letters for matrices.
Superscripts $(\cdot)^*$, $(\cdot)^{\rm T}$, $(\cdot)^{\rm H}$, $(\cdot)^{-1}$ and $(\cdot)^\dagger$ denote the conjugate, transpose, conjugate transpose, inversion and Moore-Penrose inversion operators, respectively.
$\left\| {\mathbf{A}} \right\|_F$ is the Frobenius norm of ${\mathbf{A}}$.
${\rm{vec}}( {\mathbf{A}} )$ and ${\rm{angle}}( {\mathbf{A}} )$ denote the vectorization operation and the phase values of ${\mathbf{A}}$, respectively.
${\mathbf{I}_n}$  denotes the $n\times n$ identity matrix, while $\bm{1}_n$ ($\bm{0}_n$) denotes the vector of size $n$ with all the elements being $1$ ($0$).
$\Re\{\cdot\}$ and $\Im\{\cdot\}$ denote the real part and imaginary part of the corresponding argument, respectively.
$[\mathbf{A}]_{m,n}$ denotes the $m$th-row and $n$th-column element of $\mathbf{A}$, while $[\mathbf{A}]_{[:,m:n]}$
is the sub-matrix containing the $m$th to $n$th columns of $\mathbf{A}$. The expectation is denoted by $\mathbb{E}(\cdot)$. $\frac{{\partial a}}{{\partial b}}$ is the partial derivative of $a$ with respect to $b$.

\section{System and Channel Model}\label{S2}


We investigate the multi-user downlink image transmission in near-space airship-borne massive MIMO network, where the airship-borne BS transmits image data to $K$ single-antenna user equipment (UEs). The system adopts OFDM with $N_c$ subcarriers, and the airship-borne BS is equipped with a massive antenna array with $N_t$ antennas. The transmitted red-green-blue three-dimensional (3D) image data of the $k$th UE is denoted as $\mathbf{D}[k]\in\mathbb{R}^{3\times M_x\times M_y}$, where $M_x$ and $M_y$ denote the width and height of the image, respectively. The airborne BS encodes the image data into a complex-valued symbol of dimension $N_s$ $\mathbf{s}[k]\in\mathbb{C}^{N_s\times 1}$. This coding process can be expressed as
\begin{equation}\label{eqEnC} 
	\mathbf{s}[k] = \mathcal{E} ( \mathbf{D}[k] ) ,
\end{equation}
where $\mathcal{E}(\cdot)$ denotes the encoding function that maps the image data $\mathbf{D}[k]$ onto the complex-valued vector $\mathbf{s}[k]$.
To mitigate inter-user  interference during multi-user image data transmission, the airship-borne BS employs  multi-user beamforming, converting $\mathbf{s}[k]$ for $1\leq k\leq K$ into the 3D transmit signals $\mathbf{X} \in\mathbb{C}^{N_c \times N_t\times Q}$ spanning $Q$ OFDM symbols, which can be expressed as
\begin{equation}\label{eqMIMOb} 
	\mathbf{X} = \mathcal{P}\left( \mathbf{s}[1], \cdots ,\mathbf{s}[K],\hat{\mathbf{H}} \right),
\end{equation}
where $\hat{\mathbf{H}}\! \in\! \mathbb{C}^{K\times N_c\times N_t}$ is the estimate of the true 3D CSI matrix $\mathbf{H}\! =\! \left[ [\mathbf{h}[1,1], \cdots ,\mathbf{h}[1,N_c]]; \cdots ; [\mathbf{h}[K,N_c],\cdots , \mathbf{h}[K,N_c]] \right]^{\rm T}\! =\! \Big[ \mathbf{H}^{\rm T}[1] ; \cdots ; \mathbf{H}^{\rm T}[K]\Big]^{\rm T}\! \in\! \mathbb{C}^{K\times N_c\times N_t}$, $\mathbf{h}[k,n]\in\mathbb{C}^{N_t\times 1}$ is the channel vector from the airship BS to the $k$th UE on the $n$th subcarrier, and $ \mathbf{H}[k]\! =\! \big[\mathbf{h}[k,1], \cdots ,\mathbf{h}[k,N_c]\big]^{\rm T}\!\in\! \mathbb{C}^{N_c\times N_t}$.
Since multi-user beamforming can be linear or nonlinear, we use $\mathcal{P}(\cdot)$ to represent the generic transformation from the encoded data and estimated CSI onto the corresponding transmit signals. 

The signal received by the $k$th UE on the $n$th subcarrier of the $q$th OFDM symbol can be expressed as
\begin{equation}\label{eqRx-k} 
	y[k,q,n] = \mathbf{h}^{\rm H}[k,n] \mathbf{x}[q,n] + z[k,q,n] ,
\end{equation}
where $\mathbf{x}[q,n] \in\mathbb{C}^{N_t \times 1}$ is the airship-borne BS's transmit signal on the $n$th subcarrier of the $q$th OFDM symbol, and $z[k,q,n]\sim {\cal CN}\left( 0,\sigma_n^2 \right)$ is the complex-valued additive white Gaussian noise (AWGN) with zero mean and variance $\sigma_n^2$.
By aggregating the received signals across the $N_c$ subcarriers and $Q$ OFDM symbols, the $k$th UE obtains the overall received signal ${\mathbf Y}[k] \in\mathbb{C}^{N_c \times Q}$. Based on ${\mathbf Y}[k]$, the UE decodes and reconstructs the image as 
\begin{equation}\label{eqDeC} 
	\hat{\mathbf{D}}[k] = {\cal D}\left( \mathbf{Y}[k] \right),
\end{equation}
where $\mathcal{D}(\cdot )$ denotes the decoding mapping from the received signal $\mathbf{Y}[k]$ onto the reconstructed 3D image $\hat{\mathbf{D}}[k]\! \in\! \mathbb{R}^{3\times M_x\times M_y}$.

\begin{figure*}[!t]
\vspace*{-1mm}
\centering
\includegraphics[width = 2\columnwidth,keepaspectratio]{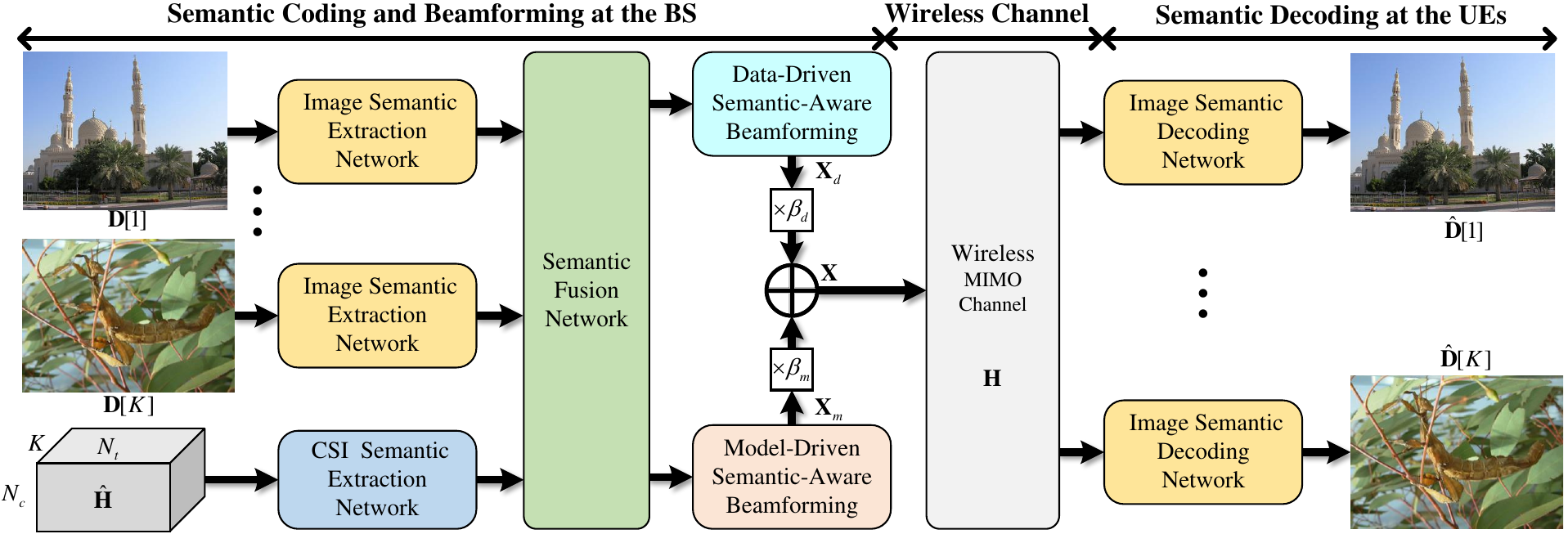}
\captionsetup{font={footnotesize}, singlelinecheck = off, justification = raggedright,name={Fig.},labelsep=period}
\caption{Proposed deep JSCBF scheme for near-space airship-borne massive MIMO image transmission network.}
\label{fig:Semantic} 
\vspace*{-4mm}
\end{figure*}


We consider a typical multipath massive MIMO channel model. Specifically, given $L_p[k]$ as the number of multipath components between the airship-borne BS and the $k$th UE, the corresponding channel vector on the $n$th subcarrier can be described as  \cite{JSAC_Wu1}
\begin{equation}\label{form4} 
	\mathbf{h}[{k,n}] = \frac{1}{\sqrt{L_p[k]}} \sum_{l=1}^{L_p[k]} \alpha_{l,k} \mathbf{a}_t\big(\theta_{l,k},\phi_{l,k}\big) e^{-\textsf{j} \frac{2\pi n\tau_{l,k}}{N_c T_s}} .
\end{equation}
In (\ref{form4}), $\alpha_{l,k} \sim \mathcal{CN}(0,1)$ is the complex gain of the $l$th path, $\theta_{l,k}\in[-\pi,\, \pi]$ and $\phi_{l,k}\in [0,\, \pi/4]$ are the $l$th path's azimuth and zenith angles of departures between the airship-borne BS and the $k$th UE, respectively, while $\tau_{l,k}$ is the delay of the $l$th path, $T_s$ is the OFDM sampling interval, and $\mathbf{a}_t(\cdot)\in\mathbb{C}^{N_t\times 1}$ denotes the  normalized transmit array response vector.

For the airship BS equipped with a half-wavelength uniform planar array (UPA) of dimension $N_t=N_y\times N_z$, the array response vector can be expressed as
\begin{align}\label{form10} 
	\mathbf{a}_t(\theta ,\phi ) = & \Big[1, \cdots ,e^{\textsf{j} \frac{2 \pi}{\lambda} d \big(n \sin(\theta ) \cos(\phi ) + m \sin(\phi )\big)}, \nonumber \\
	& \cdots , e^{\textsf{j} \frac{2 \pi}{\lambda } d \big( (N_{y} - 1) \sin(\theta ) \cos(\phi ) + (N_{z} - 1) \sin(\phi )\big)}\Big]^{\rm T},
\end{align}
where $\lambda$ is the wavelength, and the adjacent antenna spacing $d$ is given by $d = \frac{\lambda}{2}$.

\section{Deep Joint Semantic Coding and Beamforming (JSCBF)}\label{S:precoding} 

The block diagram of the proposed deep JSCBF scheme is shown in Fig.~\ref{fig:Semantic}. In this section, we first introduce the proposed deep JSCBF problem for the airship-based massive MIMO image transmission network in near space. Second, based on the transformer architecture, we design image and CSI semantic extraction networks, respectively, and further develop a semantic fusion network to fuse the extracted source semantic and CSI semantic into complex-valued semantic features for subsequent physical layer transmission. Third, we propose data-driven and model-driven semantic-aware beamforming networks to map these fused semantic features onto transmission signals. Finally, we perform joint training of all proposed networks to achieve high-quality image compression and reconstruction.

\subsection{Problem Formulation}\label{S3.1}

To achieve efficient and accurate image transmission over an airborne massive MIMO communication network with limited wireless resources, it is crucial to jointly design the image coding and beamforming at the transmitter and the image decoding at the receiver by minimizing the semantic loss between the original image and the reconstructed image. This optimization problem can be formulated as
\begin{equation}\label{equ:problem} 
	\begin{array}{cl}
		\min\limits_{{\cal E}(\cdot ),{\cal P}(\cdot ),{\cal D}(\cdot )} & \sum\limits_{k = 1}^K {\cal L}\big(\hat{\mathbf{D}}[k],\mathbf{D}[k]\big) , \\
		\text{s.t.} & \mathbf{s}[k] = {\cal E}(\mathbf{D}[k]) , \, \forall k, \\
		& \mathbf{X} = {\cal P}\left( \mathbf{s}[1], \cdots ,\mathbf{s}[K],\hat{\mathbf{H}} \right), \\
		& \hat{\mathbf{D}}[k] = \mathcal{D}(\mathbf{Y}[k]), \, \forall k, \\
		& \| \mathbf{X} \|_F^2) \le Q P_t ,
	\end{array}
\end{equation}
where the transmit signals $\mathbf{X}$ should satisfy the power constraint, and ${\cal L}\big(\hat{\mathbf{D}}[k],\mathbf{D}[k]\big)$ is the loss function to measure the semantic similarity between the reconstructed and original images. Various metrics such as Peak Signal to Noise Ratio (PSNR), MS-SSIM, or LPIPS can be used to quantify the semantic similarity between $\hat{\mathbf{D}}[k]$ and $\mathbf{D}[k]$, which will be discussed in detail in the following subsections. 

The problem (\ref{equ:problem}) is a complicated joint optimization. Conventional methods typically focus on optimizing each module separately using different metrics, thereby i) failing to achieve joint optimization based on the evaluation metric of semantic reconstruction, and ii) failing to jointly exploit the embedded semantic information in both source and CSI for designing communication systems.  As a result, conventional methods cannot achieve optimal performance, especially under imperfect CSI.
How to effectively utilize the semantic information inherent in both source and CSI to facilitate the joint design of all modules and construct an E2E massive MIMO semantic communication system for improved performance is the core problem to be solved in this paper.

\begin{figure*}[!t]
\vspace*{-1mm}
\centering
\includegraphics[width = 2\columnwidth,keepaspectratio]{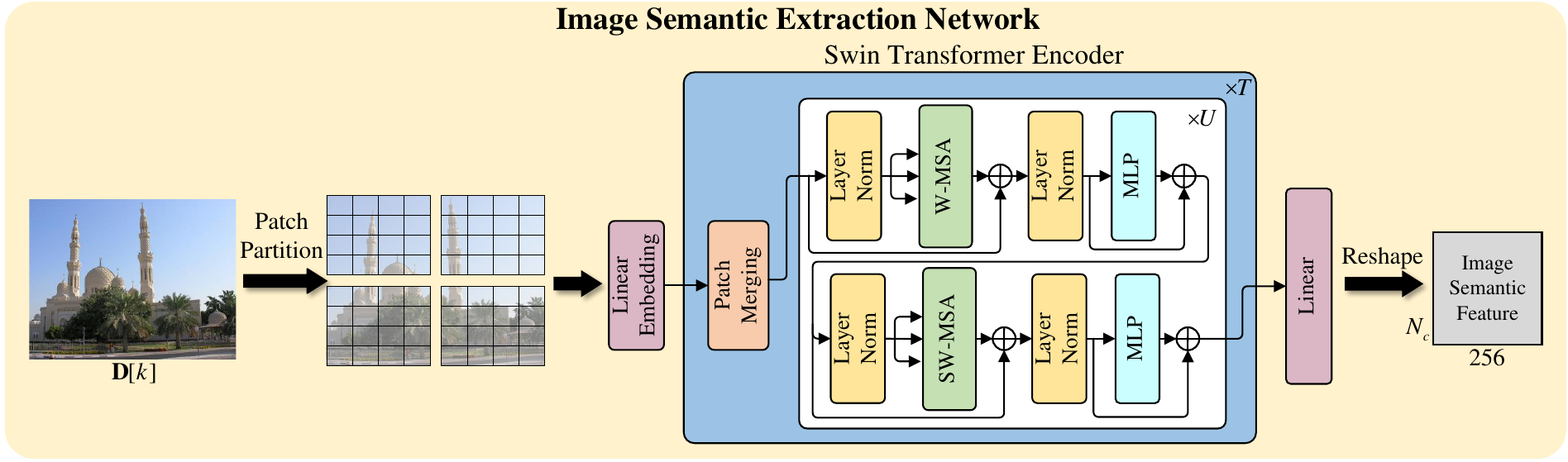}
\captionsetup{font={footnotesize}, singlelinecheck = off, justification = raggedright,name={Fig.},labelsep=period}
\vspace*{-1mm}
\caption{Image semantic extraction network, where MLP stands for multilayer perceptron, and $\times U$ ($\times T$) means the module repeats $U$ times ($T$ times).}
\label{fig:image_sem} 
\vspace*{-4mm}
\end{figure*}

\begin{figure*}[!b]
\vspace*{-4mm}
\centering
\includegraphics[width = 2 \columnwidth,keepaspectratio]{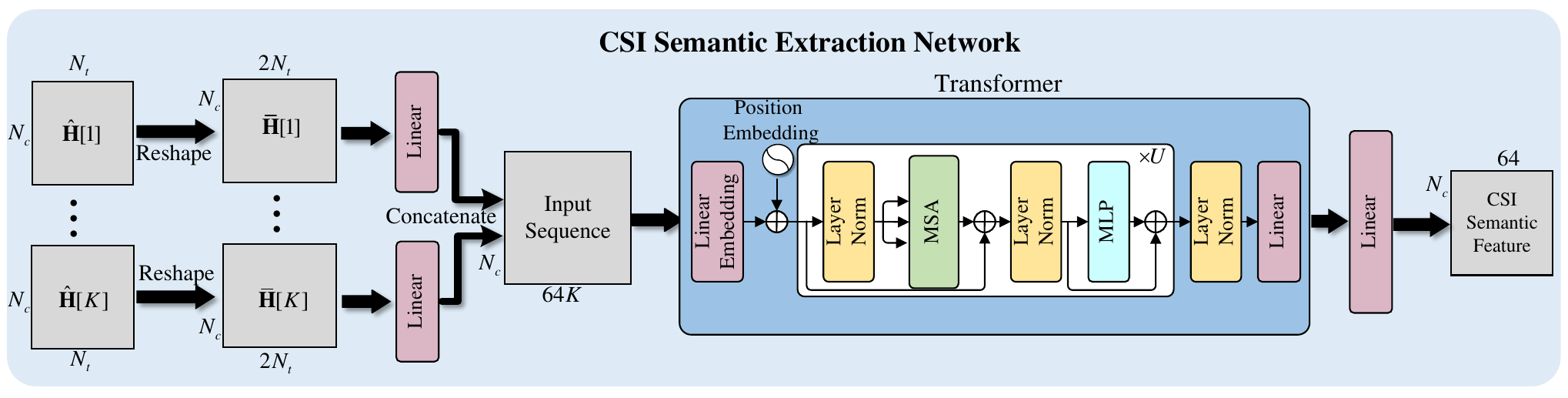}
\captionsetup{font={footnotesize}, singlelinecheck = off, justification = raggedright,name={Fig.},labelsep=period}
\vspace*{-1mm}
\caption{CSI semantic extraction network.}
\label{fig:CSI_sem} 
\end{figure*}

\subsection{Transformer-based Semantic Extraction and Fusion of Image and CSI semantics}\label{S3.2}

\subsubsection{Image semantic extraction}

As shown in Fig.~\ref{fig:image_sem}, this paper introduces a Swin Transformer-based image semantic extraction network designed for efficient image semantic extraction. The Swin Transformer has unique advantages, such as hierarchical structure processing and shifted window-based self-attention \cite{swin}.  These features facilitate detailed contextual analysis of images, making Swin Transformer an ideal network backbone for the task of image semantic extraction.

To process $3\times 256\times 256$ images with Swin Transformer, the images are first segmented into non-overlapping patches of dimension $N\times N$ by patch partitioning, which converts their dimension to $3N^2 \times\frac{256}{N}\times\frac{256}{N}$, where $N$ is the patch size. The images are further divided into $\frac{256^2}{N^2M^2}$ non-overlapping windows of dimension $M\times M$, where $M$ is the window size.
A linear embedding layer is applied to the raw patches in each window to create a ${\bf F}_p\in\mathbb{R}^{C\times M\times M}$ feature map with $C$ channels.
The feature map ${\bf F}_p$ is transformed into a sequence ${\bf F}_s\in\mathbb{R}^{M^2\times C}$, where $M^2$ is the number of patches in each window.
Next, the Swin Transformer encoder applies window-based multi-head self-attention (W-MSA) layers within each window to increase focus on local features. This attention mechanism is applied within the windows of these patches and is critical for capturing intricate details, preserving feature locality while reducing computational complexity compared to global attention mechanisms. To capture a broader context, the Swin Transformer encoder uses shifted window-based multi-head self-attention (SW-MSA) layers. This allows interaction between adjacent windows, facilitating more comprehensive global semantic feature extraction from the image.
As the image propagates through the Swin Transformer encoder layers, patch merging is applied, scaling the channel dimensions while reducing the spatial dimensions. This merging process is essential for constructing a hierarchical representation, starting with fine-grained details and progressing to more abstract features. Through patch merging, the Swin Transformer encoder ensures comprehensive and nuanced processing of visual data, deftly balancing local feature extraction with global contextual understanding.
Finally, the extracted features are passed through a fully connected linear layer, resulting in the semantic feature matrix of the image with a dimension of $N_c\times 256$.

\subsubsection{CSI semantic extraction} 

Based on the estimated CSI $\hat{\mathbf{H}}$, the airship BS extracts CSI semantic information for subsequent semantic fusion and physical layer beamforming design. This is achieved by the proposed transformer-based CSI semantic extraction network shown in Fig.~\ref{fig:CSI_sem}.
Unlike the convolution operation in CNNs \cite{csinet1}, which is limited to feature extraction from local regions, the self-attention mechanism inherent in the transformer structure excels at global feature extraction \cite{Transformer,VIT}. This capability allows it to detect the inter-subcarrier correlation within the input signal on a global scale and assign appropriate weighting coefficients to the components within each subcarrier, thereby improving overall performance. A typical transformer takes a one-dimensional (1D) real-valued sequence as input and produces an output in a similar format \cite{Transformer,VIT}. To process the complex-valued input $\hat{\mathbf{H}}[k]\! \in\! \mathbf{R}^{N_c\times N_t}$, where $\hat{\mathbf{H}}[k]$ is the estimate of $\mathbf{H}[k]$, the first step is to convert it into a real-valued matrix $\bar{\mathbf{H}}[k]\! \in\! \mathbb{R}^{N_c\times 2N_t}$:
\begin{align}\label{eqCtoR} 
	\left\{ \begin{array}{c}
		\left[ \bar{\mathbf{H}}[k] \right]_{[:,1:{N_t}]} = \Re \left\{ \hat{\mathbf{H}}[k] \right\}, \\
		\left[ \bar{\mathbf{H}}[k] \right]_{[:,1 + {N_t}:2{N_t}]} = \Im \left\{ \hat{\mathbf{H}}[k] \right\} .
	\end{array} \right.
\end{align}
As shown in Fig.~\ref{fig:CSI_sem}, the real-valued CSI matrices corresponding to all $K$ UEs are first dimensionally compressed to $N_c\times 64$ by a fully connected linear layer, and then concatenated into a 1D real-valued sequence of dimension $N_c\times 64K$. This sequence serves as the input to the transformer, where the effective input sequence length is determined by the number of subcarriers $N_c$. Inside the transformer, the input sequence is first transformed into a vector sequence of dimension $d_{\rm model}$ via a fully connected linear embedding layer followed by a position embedding layer. Different frequencies of sine functions are used to denote the positions of different subcarriers, and the position information of the subcarriers is embedded by summing with the vector sequence. The transformer then employs $U$ identical layers to extract semantic features from the input sequence, where each layer consists of a multi-head self-attention (MSA) sublayer and an MLP sublayer. The extracted features are then processed through a fully connected linear layer to obtain a CSI semantic feature matrix of dimension $N_c\ times 64$.

\begin{figure*}[!t]
\vspace*{-1mm}
\centering
\includegraphics[width = 2 \columnwidth,keepaspectratio]{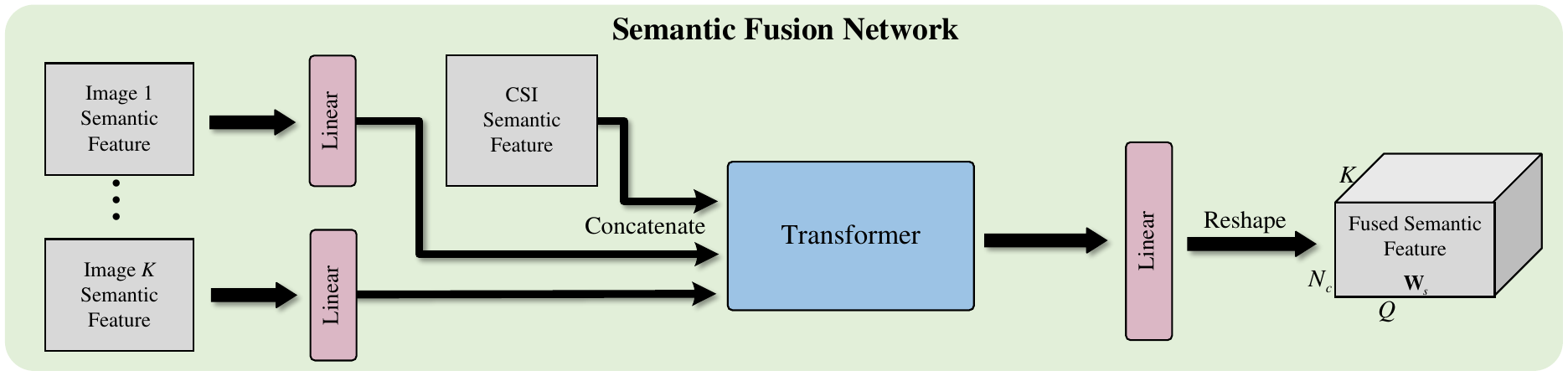}
\captionsetup{font={footnotesize}, singlelinecheck = off, justification = raggedright,name={Fig.},labelsep=period}
\vspace*{-1mm}
\caption{Semantic fusion network.}
\label{fig:sem_fuse} 
\vspace*{-4mm}
\end{figure*}

\begin{figure*}[!b]
\vspace*{-4mm}
\centering
\includegraphics[width = 2 \columnwidth,keepaspectratio]{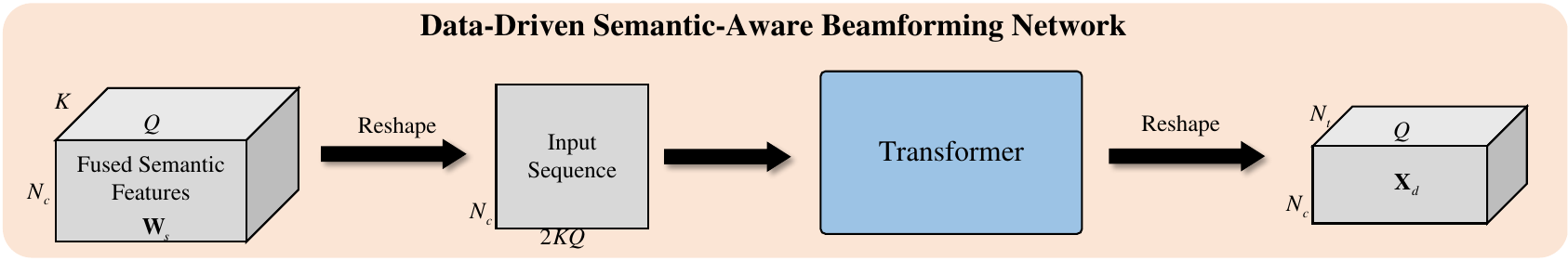}
\captionsetup{font={footnotesize}, singlelinecheck = off, justification = raggedright,name={Fig.},labelsep=period}
\vspace*{-1mm}
\caption{Data-driven semantic-aware beamforming network.}
\label{fig:DDDL} 
\vspace*{-1mm}
\end{figure*}

\subsubsection{Semantic fusion of image and CSI semantics} 

After extracting semantic features from both images and CSI, the obtained features remain in the hidden space of the neural network and cannot be directly used for physical layer transmission. To this end, we employ a semantic fusion network as shown in Fig.~\ref{fig:sem_fuse} to fuse the image and CSI semantic features from multiple UEs for semantic coding to form a semantic data stream tailored for physical layer transmission. In contrast, traditional physical layer processing schemes adopt a separate module design approach, i.e., the source coding typically considers only the distribution of the source itself, neglecting the influence of CSI. In particular, small-scale channel fading would introduce significant variations in channel conditions between different subcarriers and spatial subchannels. This can lead to degraded performance of conventional coding schemes.

The proposed approach addresses this issue by incorporating additional semantic information from the CSI at the semantic coding stage, thereby facilitating the semantic coding module to adapt to the MIMO physical layer transmission link state. This integration facilitates E2E joint optimization with the subsequent physical layer transmission modules, thereby improving the system's adaptability to channel variations. 
Specifically, we first process the image semantic feature of each UE through a linear layer and concatenate them with the CSI semantic feature. Subsequently, we the concatenated data stream using a transformer to obtain a semantic matrix of dimension $N_c \times 2QK$, denoted as $\tilde{\mathbf{W}}_{\rm{s}}$.
Then, we decompose $\tilde{\mathbf W}_{\rm s}$ to form a complex-valued matrix $\mathbf{W}_{\rm s}$ according to
\begin{equation}\label{eqFus} 
 \left\{ \begin{array}{c}
		\Re \left\{\mathbf{W}_{\rm s} \right\} = \left[ \tilde{\mathbf{W}}_{\rm{s}}\right]_{[:,1:Q K]} , \\
		\Im \left\{\mathbf{W}_{\rm s} \right\} = \left[ \tilde{\mathbf{W}}_{\rm{s}}\right]_{[:,1+Q K:2 Q K]} .
	\end{array} \right.
\end{equation}
Finally, we reshape $\mathbf{W}_{\rm s}$ into the dimension of $N_c \times K \times Q$, namely, $\mathbf{W}_{\rm s}=\big[\mathbf{W}_{\rm s}[1]; \cdots ; \mathbf{W}_{\rm s}[N_c]\big]$ with $\mathbf{W}_{\rm s}[n] \in\mathbb{C}^{K\times Q}$, for subsequent physical-layer transmission.

This proposed semantic encoding scheme leverages semantic information from both images and CSI, enabling adaptive optimization of semantic transmission based on channel conditions.

\subsection{Semantic-Aware Beamforming}\label{S3.3}

\begin{figure*}[!t]
\vspace*{-1mm}
\centering
\includegraphics[width = 2 \columnwidth,keepaspectratio]{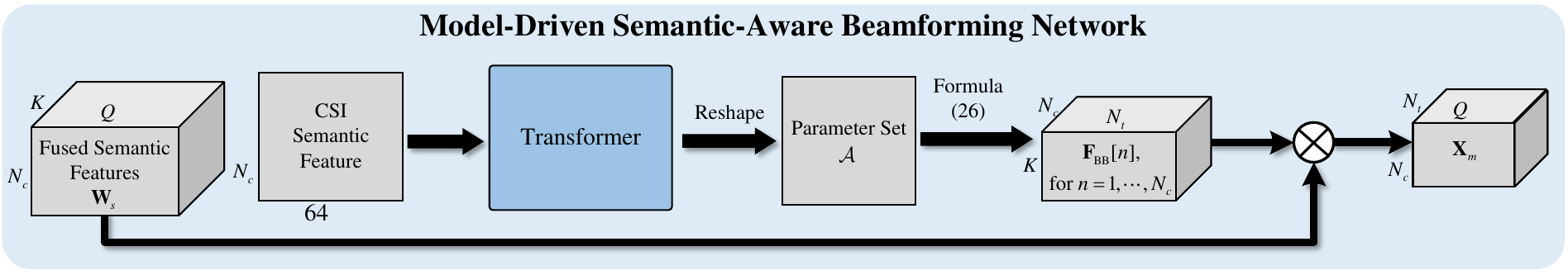}
\captionsetup{font={footnotesize}, singlelinecheck = off, justification = raggedright,name={Fig.},labelsep=period}
\vspace*{-1mm}
\caption{Model-driven semantic-aware beamforming network.}
\label{fig:MDDL} 
\vspace*{-4mm}
\end{figure*}

After the semantic fusion is completed, the semantic features are precoded in the spatial domain and transmitted to the target UE over the MIMO channel.
Since our ultimate goal is to ensure the similarity of the original and reconstructed images at the semantic level, we need to optimize the beamforming process based on the semantic information. 
To this end, we model the beamforming module as a DNN with semantic features as inputs, which is jointly trained E2E with the other modules using the final semantic metrics.
Specifically, we propose a beamforming module as a data-driven semantic-aware beamforming network, as shown in Fig.~\ref{fig:DDDL}, and a model-driven semantic-aware beamforming network, as shown in Fig.~\ref{fig:MDDL}. The outputs of these two networks are then summed to produce the final transmit signal.

Although data-driven DL approaches do not rely on expert knowledge and can adapt to imperfect CSI by training on large datasets, they suffer from slow convergence, poor interpretability, and difficulty in achieving optimal performance. In contrast, model-driven DL approaches incorporate explicit expert knowledge from the field of communications to facilitate the training of neural networks, thereby improving the network's ability to converge and achieve better performance. However, under poor channel conditions with a significant mismatch between expert knowledge and the actual system, model-driven DL solutions may fail to guide appropriate neural network design, leading to unsatisfactory performance \cite{JSAC_Wu2}. Therefore, we integrate both data-driven and model-driven DL methods for beamforming in our design to achieve improved performance.

\subsubsection{Data-driven semantic-aware beamforming}

As shown in Fig.~\ref{fig:DDDL}, our data-driven semantic-aware beamforming network maps the 3D fused semantic feature matrix $\mathbf{W}_{\rm s}\! \in\! \mathbb{C}^{N_c\times K\times Q}$ onto the 3D transmit signal matrix $\mathbf{X}_d\! \in\! \mathbb{C}^{N_c\times N_t\times Q}$. 
Specifically, the complex-valued semantic feature matrix $\mathbf{W}_{\rm s}$ is first transformed into a 1D real-valued input sequence with dimension $N_c\times 2QK$, then processed by a transformer for further refinement, and finally transformed into the transmit signals $\mathbf{X}_d$ by a fully connected linear layer and subsequent transform operation.

Current massive MIMO beamforming primarily uses linear beamforming, and different linear beamforming strategies are used for different scenarios. For example, space division multiple access beamforming schemes perform well in the case of sufficiently accurate CSI and small channel correlation between UEs, while non-orthogonal multiple access and rate-split multiple access beamforming techniques show superior performance in the case of imperfect CSI or severe channel interference between UEs \cite{JSAC_Wu2}. In contrast, we implement beamforming through the nonlinear mapping provided by DNN, and exploit the powerful representational capability of DNN to provide the proposed beamforming scheme with enhanced adaptability.

\subsubsection{Model-driven semantic-aware beamforming}

We initially derive a novel WMMSE beamforming method tailored for multi-user MIMO scenarios under imperfect CSI. Subsequently, we incorporate DL techniques into  training to enhance performance, culminating in the establishment of a model-driven semantic-aware beamforming network. 

\emph{2.1)}~Considering the utilization of linear beamforming based on WMMSE with sum rate as the optimization objective, the optimization problem can be articulated as
\begin{align}\label{equ:model_opt1} 
  \begin{array}{cl}
	  \max\limits_{\mathbf{F}_{\rm BB}[n],\forall n} & R = \frac{1}{N_{c}} \sum\limits_{k=1}^K \sum\limits_{n=1}^{N_c} \\
	  & \log_{2}\left( 1 + \frac{\big| \mathbf{h}^{\rm H}[k,n] \mathbf{f}_{\rm BB}[k,n] \big|^{2}}{\sum\limits_{m\neq k}\big| \mathbf{h}^{\rm H}[k,n] \mathbf{f}_{\rm BB}[m,n] \big|^{2} + \sigma_n^2} \right) , \\
	 \text{s.t.} &  \sum\limits_{n = 1}^{N_c} \left\| \mathbf{F}_{\rm BB}[n] \right\|_F^2 \le P_t ,
	\end{array}
\end{align}
where $\mathbf{f}_{\rm BB}[k,n]\! \in\! \mathbb{C}^{N_t\times 1}$ represents the beamforming vector for the $k$th UE on the $n$th subcarrier, $\mathbf{F}_{\rm BB}[n]=\big[ \mathbf{f}_{\rm BB}[1,n],\cdots, \mathbf{f}_{\rm BB}[K,n]\big]\!\in\! \mathbb{C}^{N_t\times K}$, and the beamforming matrices $\mathbf{F}_{\rm BB}[n]$, $\forall n$, are required to satisfy a power constraint with a maximum power of $P_t$. 

In the optimization problem (\ref{equ:model_opt1}), the power allocation in the frequency domain and the design of beamforming in the spatial domain are intricately coupled, rendering the resolution of this optimization problem a challenging task. Consequently, we employ the concept of alternating optimization, decoupling the power allocation in the frequency domain and the beamforming in the spatial domain, and optimizing them in an alternating fashion. Specifically, reformulating (\ref{equ:model_opt1}) as
\begin{align}\label{equ:model_opt2} 
  \begin{array}{cl}
	  \max\limits_{\mathbf{F}_{\rm BB}[n],p[n],\forall n} & R = \frac{1}{N_{c}} \sum\limits_{k=1}^K \sum\limits_{n=1}^{N_c} \\
	  & \log_{2}\left( 1 + \frac{\big| \mathbf{h}^{\rm H}[k,n] \mathbf{f}_{\rm BB}[k,n] \big|^{2}}{\sum\limits_{m\neq k}\big| \mathbf{h}^{\rm H}[k,n] \mathbf{f}_{\rm BB}[m,n] \big|^{2} + \sigma_n^2} \right) , \\
	 \text{s.t.} & \left\| \mathbf{F}_{\rm BB}[n] \right\|_F^2 \le p[n] , \, \forall n , \\
	 & \sum\limits_{n = 1}^{N_c} p[n]  \le P_t,
	\end{array}
\end{align}
where $p[n]$ denotes the power allocation coefficient on the $n$th subcarrier. Given the fixed beamforming matrices $\mathbf{F}_{\rm BB}[n],\forall n$, we can optimize the power allocation coefficients $p[n],\forall n$ in the frequency domain, utilizing the water-filling algorithm. Given the power allocation coefficients, the optimization of the beamforming matrices can be decoupled on per subcarrier base as
\begin{align}\label{equ:model_opt3} 
  \begin{array}{cl}
	  \max\limits_{\mathbf{F}_{\rm BB}[n]} & R[n] = \sum\limits_{k=1}^K \\
		& \log_{2}\left( 1 + \frac{\big| \mathbf{h}^{\rm H}[k,n] \mathbf{f}_{\rm BB}[k,n] \big|^{2}}{\sum\limits_{m\neq k}\big| \mathbf{h}^{\rm H}[k,n] \mathbf{f}_{\rm BB}[m,n] \big|^{2} + \sigma_n^2} \right) , \\
	  \text{s.t.} &  \left\| \mathbf{F}_{\rm BB}[n] \right\|_F^2 \le p[n] .
	\end{array}
\end{align}
That is, given $p[n],\forall n$ , the beamforming matrices for different subcarriers can be designed independently. 

Consider the relationship between the true CSI $\mathbf{h}[k,n]$ and the estimated CSI $\hat{\mathbf{h}}[k,n]$ as
\begin{equation}\label{eqCSIerror} 
	\mathbf{h}[k,n] = \hat{\mathbf{h}}[k,n] + \Delta\mathbf{h}[k,n],
\end{equation}
where $\Delta \mathbf{h}[k,n]\! \in\! \mathbb{C}^{N_t\times 1}$ represents the CSI error, which follows the complex Gaussian distribution with the auto-correlation matrix $\mathbf{R}_e\in\mathbb{C}^{N_t\times N_t}$. Denote
\begin{align}\label{eqEqS} 
  \hat{r}[k,n] =& e[k,n] \big( \mathbf{h}^{\rm H}[k,n] \mathbf{F}_{\rm BB}[n] \mathbf{r}[n] + z[k,n]\big)
\end{align}
as the estimate of the transmit data $r[k,n]$ to the $k$th UE on the $n$th subcarrier, where $\mathbf{r}[n]\! =\! \big[r[1,n],\cdots ,r[K,n]\big]^{\rm T}\!\in\! \mathbf{C}^{K\times 1}$ is the transmit data vector on the $n$th subcarrier, $e[k,n]$ is the equalizer of the $k$th UE on the $n$th subcarrier, and $z[k,n]\!\sim\! {\cal CN}(0,\sigma^2_n)$ is the complex-valued AWGN. Then the MSE of decoding the data for the $k$th UE can be approximately expressed as
\begin{equation}\label{equ:eps} 
	\varepsilon [k,n] = \mathbb{E} \left\{ \left| \hat{r}[k,n] - r[k,n] \right|^2 \right\} \approx \varepsilon^{(1)}[k,n] + \varepsilon^{(2)}[k,n],
\end{equation}
where
\begin{equation}\label{equ:eps1} 
  \varepsilon^{(1)}[k,n] = |e[k,n]|^2 T[k,n] - 2\Re \left\{ e[k,n] \hat{\mathbf{h}}^{\rm H}[k,n] \mathbf{f}_{\rm BB}[k,n] \right\} + 1
\end{equation}
represents the MSE under the assumption that the CSI error is negligible, and
\begin{equation}\label{equ:eps2} 
  \varepsilon^{(2)}[k,n] = |e[k,n]|^2 \sum\limits_{m \neq k} \mathbf{f}_{\rm BB}^{\rm H}[m,n] \mathbf{R}_e \mathbf{f}_{\rm BB}[m,n]
\end{equation}
represents the MSE introduced by the inter-user interference due to the CSI error, while
\begin{equation}\label{equ:eps3} 
  T[k,n] = \sum\limits_{m = 1}^K \left| \hat{\mathbf{h}}^{\rm H}[k,n] \mathbf{f}_{\rm BB}[m,n] \right|^2 + \sigma _n^2
\end{equation}
denotes the average received power.

We now elaborate the formula (\ref{equ:eps}).
Since the airship BS lacks perfect CSI, the MSE at the beamforming stage is a random variable to the airship BS that does not facilitate the calculation of the actual achievable rates. Therefore, we propose to optimize using an approximate MSE in lieu of the actual MSE. We assume the availability of equivalent CSI at the UE side post-linear beamforming, enabling the realization of the rates derived below. Additionally, we presume that the MSE is exclusively attributed to inter-user interference and noise.
For the $k$th UE, the expected received signal is $e[k,n]\big( \Delta \mathbf{h}[k,n] + \hat{\mathbf{h}}[k,n] \big)^{\rm H} \mathbf{f}_{\rm BB}[k,n] r[k,n]$, while the inter-user interference and noise are $\sum\limits_{m\neq k} e[k,n]\big( \Delta \mathbf{h}[k,n] + \hat{\bf{h}}[k,n] \big)^{\rm H} \mathbf{f}_{\rm BB}[m,n] r[m,n] + e[k,n] z[k,n]$. Based on the above reasoning, the MSE can be represented by (\ref{equ:eps}). It is important to note that we have made a simplistic assumption that the CSI error $\Delta \mathbf{h}[k,n]$ follows a complex Gaussian distribution and is independent of the true CSI $\mathbf{h}[k,n]$, hence the use of the approximation symbol in (\ref{equ:eps}).

\begin{figure*}[!b]
\vspace*{-5mm}
	\centering
	\includegraphics[width = 2 \columnwidth,keepaspectratio]{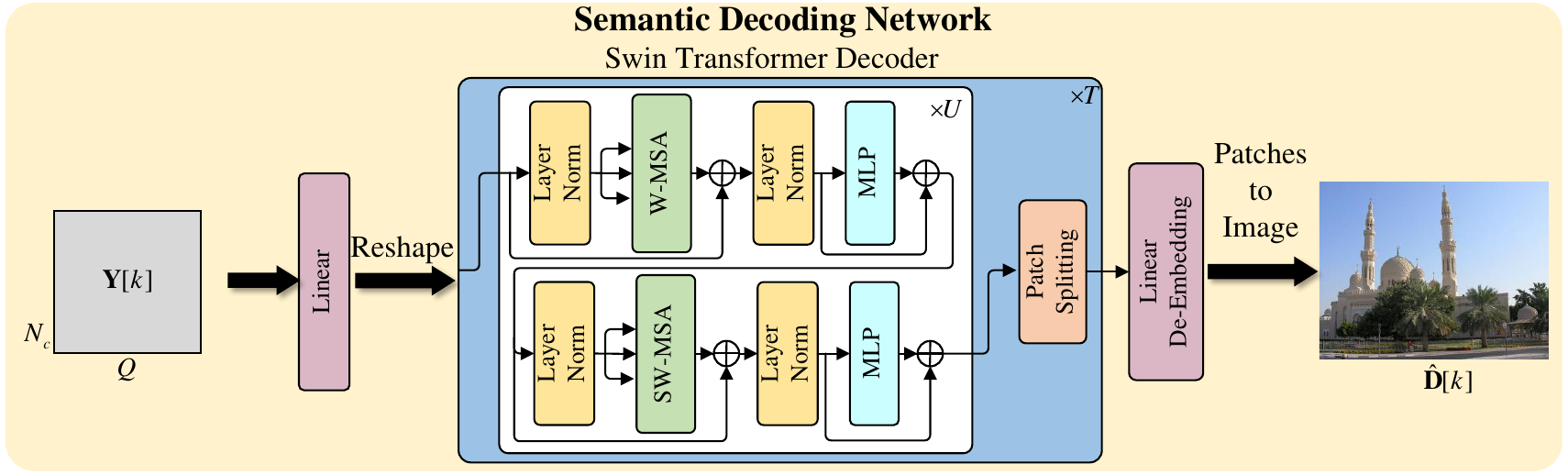}
	\captionsetup{font={footnotesize}, singlelinecheck = off, justification = raggedright,name={Fig.},labelsep=period}
	\vspace*{-1mm}
	\caption{Semantic decoding network.}
	\label{fig:seman_dec} 
	\vspace*{-1mm}
\end{figure*}

By setting $\frac{\partial \varepsilon[k,n]}{\partial e[k,n]}=0$, the minimum mean square error (MMSE) equalizer can be obtained as:
\begin{equation}\label{equ:e_MMSE} 
	e^{\rm MMSE}[k,n]\! =\! \frac{\mathbf{f}_{\rm BB}^{\rm H}[k,n] \hat{\mathbf{h}}[k,n]}{\left(\! T[k,n]\! + \!\sum\limits_{m \neq k}^K\! \mathbf{f}_{\rm BB}^{\rm H}[m,n] \mathbf{R}_e \mathbf{f}_{\rm BB}[m,n] \!\right)} .\!
\end{equation}
Substituting (\ref{equ:e_MMSE}) into (\ref{equ:eps}), the MMSE can be obtained as
\begin{equation}\label{equ:e_MMSE1} 
	\varepsilon^{\rm MMSE}[k,n] = 1 - \frac{\left| \mathbf{f}_{\rm BB}^{\rm H}[k,n] \hat{\mathbf{h}}[k,n]  \right|^2}{\left( T[k,n] + \sum\limits_{m \ne k} \mathbf{f}_{\rm BB}^{\rm H}[m,n] \mathbf{R}_e \mathbf{f}_{\rm BB}[m,n] \right)}.
\end{equation}
Then, the signal-to-interference-plus-noise ratio (SINR) for the $k$th UE on the $n$th subcarrier can be expressed as
\begin{align}\label{eqSINR} 
  \gamma[k,n] =& \frac{1}{\varepsilon^{\operatorname{MMSE}}[k,n]} - 1 ,
\end{align}
with the corresponding achievable rate given by 
\begin{align}\label{eqC[k,n]} 
  \hat{R}[k,n] =& - \log_2\left( \varepsilon^{\rm MMSE}[k,n] \right) .
\end{align} 
Since the logarithmic rate-MSE relationship cannot be directly used for solving rate optimization problems, we introduce the augmented weighted MSE (WMSE)
\begin{equation}\label{equ:xi} 
	\xi [k,n] = \lambda [k,n] \varepsilon [k,n] - \log_2(\lambda [k,n]),
\end{equation}
where $\lambda [k,n]$ represents the weight of MSE for the $k$th UE on the $n$th subcarrier. By setting $\frac{\partial \xi[k,n]}{\partial \lambda[k,n]}=0$, the optimal weight can be obtained as
\begin{equation} \label{equ:lambda_MMSE} 
	\lambda^{\rm MMSE}[k,n] = \frac{1}{\varepsilon^{\rm MMSE}[k,n]} .
\end{equation}

Substituting  (\ref{equ:e_MMSE}) and (\ref{equ:lambda_MMSE}) into (\ref{equ:xi}), the relationship for the rate-WMMSE can be established as
\begin{equation}\label{eqR-WMMS} 
	\begin{array}{cl}
		\min\limits_{\mathbf{F}_{\rm BB}[n],\mathbf{e}[n],\bm{\lambda}[n]} & \xi [n] = \sum\limits_{k=1}^K \xi [k,n] , \\
		\text{s.t.} & \left\| \mathbf{F}_{\rm BB}[n] \right\|_F^2 \le p[n] ,
	\end{array}
\end{equation}
where $\mathbf{e}[n]\! =\!\left[e[1,n], \cdots, e[K,n]\right]^{\rm T}\!\in\! \mathbb{C}^{K\times 1}$ and $\bm{\lambda}[n]\! =\! \left[\lambda[1,n], \cdots, \lambda[K,n]\right]^{\rm T}\!\in\! \mathbb{R}^{K\times 1}$. By fixing $\mathbf{e}[n]$ and $\bm{\lambda}[n]$ and applying Lemma~1 of \cite{DL_unfloding1} to eliminate the power constraints, we obtain the closed-form solution for $\mathbf{F}_{\rm BB}[n]$ from $\frac{\partial \xi [n]}{\partial \mathbf{f}_{\rm BB}[k,n]}\! =\! \mathbf{0}_{N_t}$, i.e.,
\begin{align}\label{equ:MDDL} 
	\mathbf{f}_{\rm BB}[k,n] =& \bigg( \frac{\sigma_n^2}{p[n]} \sum\limits_{m=1}^K \lambda [m,n] \left| e[m,n] \right|^2 \mathbf{I}_{N_t} \nonumber \\
	& + \sum\limits_{m \ne k} \lambda [m,n] \left| e[m,n] \right|^2 \mathbf{R}_e \nonumber \\
	& + \sum\limits_{m=1}^K \lambda [m,n] \left| e[m,n] \right|^2 \mathbf{h}[m,n] \mathbf{h}^{\rm H}[m,n] \bigg)^{-1} \nonumber \\
	&  \times e^{*}[k,n]\lambda [k,n] \mathbf{h}[k,n].
\end{align}

The beamforming matrices ${\bf F}_{\rm BB}[n], \forall n$, depend on the parameter set $\mathcal{A}\! = \! \big\{ p[n],\lambda [k,n],e[k,n],\mathbf{R}_e; \, \text{for}\, k\! =\! 1, \cdots ,K;n\! =\! 1, \cdots ,{N_c}\big\}$. Hence, the beamforming process can be decomposed into the block coordinate descent iterative optimization of the power allocation coefficient $p[n]$, weighting factor $\lambda[k,n]$, equalizer coefficient $e[k,n]$ and beamformer $\mathbf{f}_{\rm BB}[k,n]$, as summarized in Algorithm~\ref{Alg1}, where $I_1$ is the number of iterations.

\emph{2.2)}~However, the above WMMSE algorithm is based on the assumption that the CSI error follow a known complex Gaussian distribution and are independent of the CSI, which may not be true in the actual scenario. To this end, we propose a model-driven semantic-aware beamforming network by deep unfolding the proposed WMMSE algorithm, as depicted in Fig.~\ref{fig:MDDL}. This network is capable of perceiving the semantic information from the input and adaptively adjust the critical parameter set $\mathcal{A}$ through DL training for improved performance.
Specifically, this network takes the CSI semantic feature as input. A transformer is employed to process the CSI semantic feature and generate the critical parameter set $\mathcal{A}$ as outlined in the proposed WMMSE algorithm. The closed-form solution (\ref{equ:MDDL}) is then utilized to obtain the beamformer based on the critical parameter set $\mathcal{A}$. 

Based on the obtained beamformer, we perform linear beamforming on the fused semantic features to obtain the 3D transmit signal matrix $\mathbf{X}_m=\big[\mathbf{X}_m[1];\cdots ;\mathbf{X}_m[N_c]\big]\! \in\! \mathcal{C}^{N_c\times N_t\times Q}$ with
\begin{equation}\label{eqDL-B} 
	\mathbf{X}_m[n] = \mathbf{F}_{\rm BB}[n] \mathbf{W}_{\rm s}[n]\in \mathbb{C}^{N_t\times Q} ,\, \text{for}\, n = 1, \cdots ,N_c .
\end{equation}

\emph{2.3)}~To leverage the advantages of both data-driven DL and model-driven DL, we adopt a weighted summation method to accomplish the merging of two types of beamforming, yielding the 3D transmit signals
\begin{equation}\label{eqFmTB} 
	\mathbf{X}_c = \beta_d \mathbf{X}_d + \beta_m \mathbf{X}_m ,
\end{equation}
where $\beta_d$ and $\beta_m$ are the learnable weighting coefficients for weighting the two beamformers, respectively.
Through the E2E DL training, the network can strike a balance between data-driven DL and model-driven DL for achieving the optimal beamforming performance.

To ensure that the average power does not exceed $P_t$, power constraints are applied to yield the final 3D transmit signals as
\begin{equation}\label{eqFmTBpc} 
	\mathbf{X} = \min \left\{ \sqrt{Q P_t}, \| \mathbf{X}_c \|_F \right\} \frac{\mathbf{X}_c}{\| \mathbf{X}_c \|_F} .
\end{equation}

\SetAlgoNoLine
\SetAlCapFnt{\normalsize}
\SetAlCapNameFnt{\normalsize}\
\begin{algorithm}[!t]
	\caption{WMMSE-Based Beamforming with Imperfect CSI}
  \label{Alg1}
	\begin{algorithmic}[1]
		\STATE \textbf{Initialize} Set beamformer $\mathbf{F}_{\rm BB}$ to zero forcing beamformer;
		\FOR {$i=1$ to $I_1$}
		  \STATE \textbf{Update} $p[n]$ using water-filling algorithm with fixed $\mathbf{F}_{\rm BB}[n]$, $\mathbf{e}[n]$ and $\bm{\lambda}[n]$, for $n=1,\cdots,N_c$;
		  \STATE \textbf{Update} $\mathbf{e}[n]$ and $\bm{\lambda}[n]$ using (\ref{equ:e_MMSE}) and (\ref{equ:lambda_MMSE}) with fixed $\mathbf{F}_{\rm BB}[n]$ and $p[n]$, for $n=1,\cdots,N_c$;
		  \STATE \textbf{Update} $\mathbf{F}_{\rm BB}[n]$ using (\ref{equ:e_MMSE}) with fixed $\mathbf{e}[n]$, $\bm{\lambda}[n]$ and $p[n]$, for $n=1,\cdots,N_c$;
		\ENDFOR
	\end{algorithmic}
\end{algorithm}

\subsubsection{Semantic Decoding}

At the receiver of each UE, a semantic decoding network based on the Swin Transformer is used to reconstruct images from the received signals, as shown in Fig.~\ref{fig:seman_dec}. Specifically, each UE first processes the received signal through a fully connected layer and transforms it into a semantic feature map of dimension $N_c\times 16\times 16$. Then, the UE uses a Swin Transformer decoder that is symmetric with the Swin Transformer encoder to process the feature map. Unlike patch merging in the Swin Transformer encoder, the Swin Transformer decoder uses patch splitting to upsample the feature map resolution and reduce the number of feature map channels. A final linear deembedding layer is used to map the feature map onto the reconstructed image.

\subsubsection{Loss Function}

To enhance the precision of image reconstruction and improve the visual quality, we propose  a composite loss function that combines the MSE, MS-SSIM \cite{SSIM}, and LPIPS \cite{LPIPS} based on pre-trained VGG-16 \cite{VGG}, denoted as $\mathcal{L}_{\rm MSE}(\cdot)$, $\mathcal{L}_{\rm MS-SSIM}(\cdot)$, and $\mathcal{L}_{\rm VGG}(\cdot)$, respectively.

The MSE loss is utilized to minimize the average squared difference between the estimated values and the ground truth, ensuring pixel-level accuracy. 

By contrast, the MS-SSIM loss is employed to preserve the structural information across multiple scales, thereby preserving the perceptually relevant parts of the image \cite{SSIM}. 

Although MSE and MS-SSIM are the most widely used metrics for image similarity measurement, they are simplistic functions that fail to account for many nuances of human perception \cite{LPIPS}. To better achieve semantic communication, we further adopt the emerging DL-based LPIPS metric \cite{LPIPS} as a perceptual loss to quantify image transmission performance. This metric is capable of emulating the human perceptual evaluation process, thus providing an LPIPS loss score. The LPIPS is computed using features extracted from a neural network, typically a pre-trained VGG network. The LPIPS value ranges from 0 to 1, with a value closer to 0 indicating less distortion. More specifically, the LPIPS metric can be expressed as
\begin{equation}\label{eqLPIPS} 
	\mathcal{L}_{\rm LPIPS}\big(\hat{\mathbf{D}},\mathbf{D}\big) = \sum\limits_{j} w_j \left\| \phi_j\big(\hat{\mathbf{D}}\big) - \phi_j\big(\mathbf{D}\big) \right\|_2^2,
\end{equation}
where $\phi_j\big(\hat{\mathbf{D}}\big)$ and $\phi_j\big(\mathbf{D}\big)$ denote the feature maps extracted from the generated image $\hat{\mathbf{D}}$ and the target image $\mathbf{D}$ by the $j$th layer of the neural network, respectively, and $w_j$ represents the learned weight for the $j$-th layer, emphasizing its importance in the perceptual similarity measure. The selection of layers and the calculation of weights are typically based on empirical evaluations consistent with studies of human perception.

The overall loss function is a weighted sum of these three components:
\begin{align}\label{eqTloss} 
	\mathcal{L}_{\rm total}\big(\hat{\mathbf{D}},\mathbf{D}\big) =& \lambda_{\rm MSE} \mathcal{L}_{\rm MSE}\big(\hat{\mathbf{D}},\mathbf{D}\big)  \nonumber \\
	& + \lambda_{{\rm MS}-{\rm SSIM}} \mathcal{L}_{{\rm MS}-{\rm SSIM}}\big(\hat{\mathbf{D}},\mathbf{D}\big) \nonumber \\
	& + \lambda_{\rm LPIPS} \mathcal{L}_{\rm LPIPS}\big(\hat{\mathbf{D}},\mathbf{D}\big) ,
\end{align}
where $\lambda_{\rm MSE}$, $\lambda_{{\rm MS}-{\rm SSIM}}$ and $\lambda_{\rm LPIPS}$ are the weights that balance the contribution of each loss component. By combining these loss functions, our model is trained to achieve pixel-level accuracy, capture multi-scale structural similarity and maintain high-level perceptual qualities, thereby generating images that are appealing to the human visual system.

\section{Numerical Results}\label{S6}

\subsection{Baseline Schemes}\label{S6.1}

For the performance  evaluation, we compare our proposed scheme with the following baseline schemes.

\begin{itemize}
\item {{{\textbf{BPG/bmshj2018}} + {\textbf{LDPC}} + {\textbf{QAM}} + {\textbf{RZF}}}: In this case, the airship-borne BS employs the advanced image coding method better portable graphics (BPG) \cite{BPG} or the DL-based bmshj2018 model \cite{bmshj2018} for image source coding and decoding, while low-density parity-check (LDPC) code with 1/2, 2/3, or 3/4 code rate is utilized for channel coding and decoding. Constellation modulation includes binary phase shift keying (BPSK), quadrature phase shift keying (QPSK), or 16 quadrature amplitude modulation (QAM), and regularized zero forcing (RZF) is employed for MIMO beamforming.  The search for optimal combinations of source coding compression rate, LDPC code rate, and constellation modulation order is conducted to obtain the best performance  configuration for this baseline scheme.}

\item {{{\textbf{BPG/bmshj2018}} + \textbf{E2E transmission network}}: In this case, the airship BS adopts BPG or bmshj2018 model to complete the image source coding and decoding, LDPC to complete the channel coding and decoding, and extends the autoencoder-based E2E OFDM transmission network in \cite{E2Ebit} to a multi-user MIMO system for achieving  the physical-layer transmission, where the airship BS transmits 2 bits to each UE on each subcarrier during  each OFDM symbol.}
	
\item {{{\textbf{BPG/bmshj2018}} +\textbf{capacity}}: In this case, the airship BS adopts the BPG or bmshj2018 model for image source coding and decoding, and is able to transmit the coded bits perfectly at the rate of the  channel capacity. These two schemes can be considered as the ideal performance upper bounds. }
	
\item {{{\textbf{Deep JSCC}}}: In this case, we adopt the CNN-based deep JSCC scheme \cite{Deep-JSCC} for the E2E transmission of images. Notably, the original AWGN channel is substituted with an equivalent channel after RZF beamforming.}
\end{itemize}

\begin{table}[!t]
	\centering %
	\color{black}
	\captionsetup{font={color = {black}}}
	\caption{NMSE performance of GMMV-LAMP at $\text{SNR}=20$\,dB.}
	\label{table1} 
	\begin{tabular}{|l|l|l|l|l|l|}
		\hline
		\begin{tabular}[c]{@{}l@{}}Number of pilot \\ OFDM symbols $L$\end{tabular} & 4     & 8     & 16    & 32    & 64    \\ \hline
		NMSE                                                                        & 0.205 & 0.063 & 0.020 & 0.011 & 0.006 \\ \hline
	\end{tabular}
	\vspace{-4mm}
\end{table}

\subsection{Dataset for Neutral Network Training}\label{S6.2}

The channel dataset is generated following the sparse multipath channel model for airship-borne massive MIMO communication scenarios, as outlined in Subsection~\ref{S2:channel}. Specifically, the airship-borne BS is equipped with an $8\times 8$ UPA maintaining a half-wavelength antenna spacing, and the number of subcarriers is $N_c = 64$. The complex gains of different  channel paths follow an independent and identically distributed (i.i.d.) complex Gaussian distribution, i.e., $\alpha_{l,k} \sim \mathcal{CN}(0,1)$. 
The number of paths and UEs are set to 2 and 4, respectively.
Furthermore, throughout the process of training and validating the proposed deep JSCBF scheme, we consistently employ an advanced channel estimation method based on generalized multiple measurement-vectors
(GMMV)-learnable approximate message passing (LAMP) \cite{GMMV-LAMP} to acquire the estimated CSI as the input of the deep JSCBF network. This ensures that the neural network can learn to mitigate the performance degradation caused by CSI estimation errors. 
Table~\ref{table1} shows the performance of the GMMV-LAMP channel estimation scheme in terms of Normalized Mean Square Error (NMSE) at 20 dB SNR, evaluated over different numbers of OFDM pilot symbols. It can be observed that the channel estimation results are very poor in the case of low pilot overhead.  For the proposed deep JSCBF scheme, as well as for all baseline schemes, a fair comparison is made based on the imperfect CSI estimated by GMMV-LAMP.

For the image dataset, we adopt the open-source ImageNet dataset \cite{ImageNet}, which is divided into training, validation, and test sets containing 100000, 10000, and 10000 image samples, respectively.

\begin{figure*}[!t]
\vspace{-4mm}
	\centering
	\subfigure[]{
		\begin{minipage}[t]{0.33\linewidth}
			\centering
			\label{fig:simu1-a}
			\includegraphics[width = 1\columnwidth,keepaspectratio]{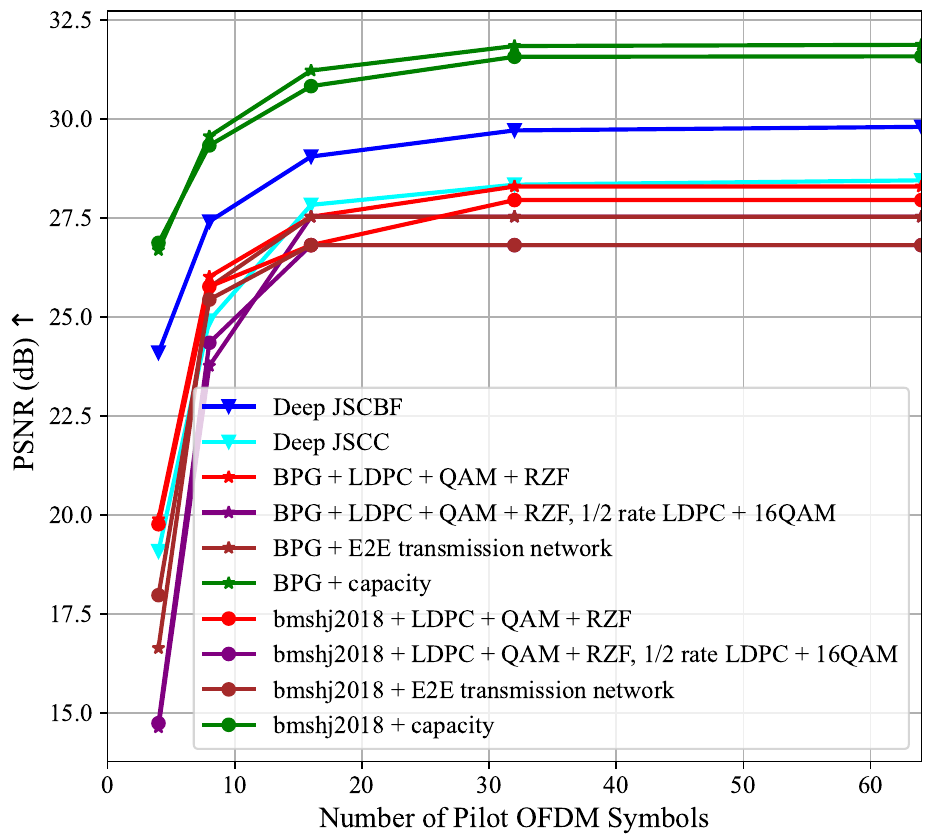}\\
		\end{minipage}%
	}%
	\subfigure[]{
		\begin{minipage}[t]{0.33\linewidth}
			\centering
			\label{fig:simu1-b}
			\includegraphics[width = 1\columnwidth,keepaspectratio]{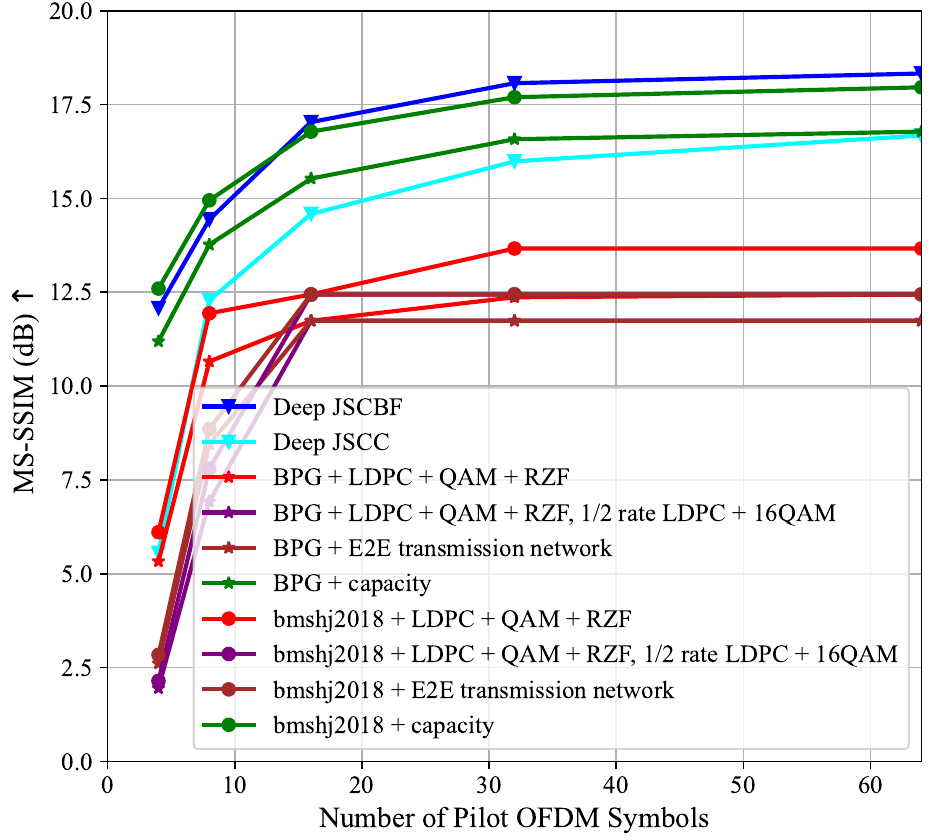}\\
		\end{minipage}%
	}%
	\subfigure[]{
		\begin{minipage}[t]{0.33\linewidth}
			\centering
			\label{fig:simu1-c}
			\includegraphics[width = 1\columnwidth,keepaspectratio]{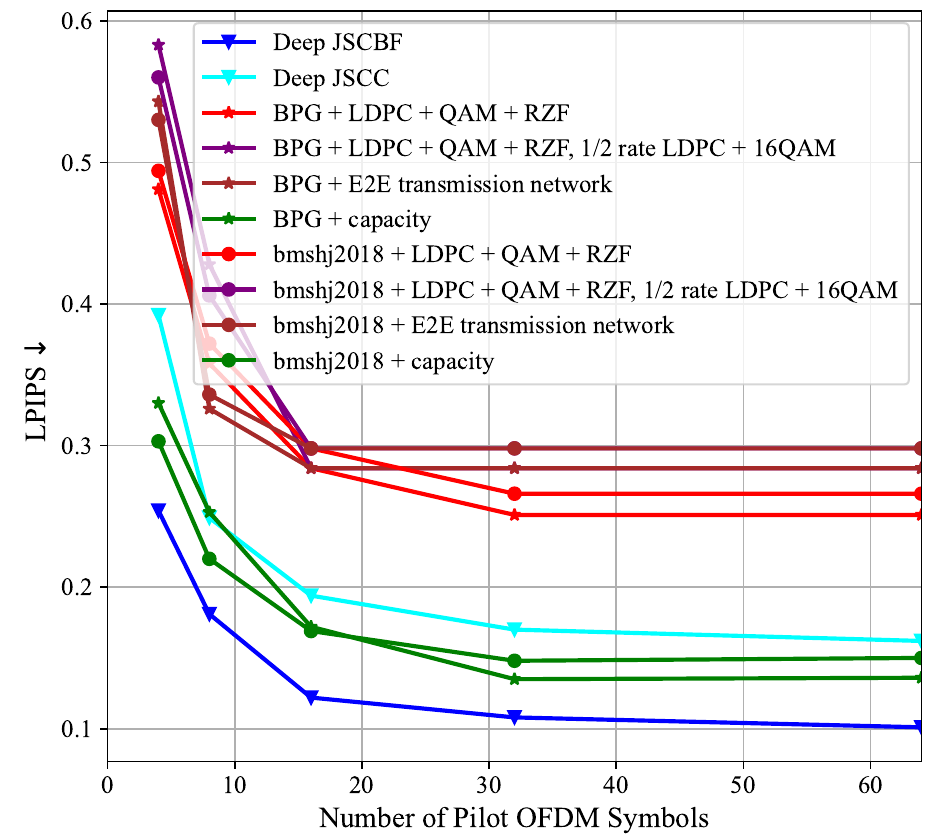}\\
		\end{minipage}%
	}%
	\centering
	\setlength{\abovecaptionskip}{-1mm}
	\captionsetup{font={footnotesize}, singlelinecheck = off, justification = raggedright,name={Fig.},labelsep=period}
\caption{Performance comparison of different solutions versus the number of pilot OFDM symbols $L$ at SNR~$=$~20 dB, given $Q = 128$: (a) PSNR performance comparison; (b) MS-SSIM performance comparison; (c) LPIPS performance comparison.}
	\label{fig:simu1} 
\vspace*{-6mm}
\end{figure*}

\subsection{Training Settings}\label{S6.3}

We use the open-source DL library PyTorch to train and validate the proposed deep JSCBF scheme on a computer with dual Nvidia GeForce GTX 3090 GPUs. During the training process, we adopt the Adam optimizer and use the loss function introduced in the previous section to train the entire neural network, with the training set batch size set to 16. The initial learning rate is set to $10^{-4}$, and a cosine annealing strategy is used to reduce the learning rate during the training process to further improve performance in the later stages of training. 
To speed up the training process, we first pre-train both the image semantic extraction network and the semantic decoding network without considering the physical layer transmission process. Then, these networks are integrated with other modules for joint E2E training.

\begin{figure*}[!b]
	\vspace{-5mm}
	\centering
	\subfigure[]{
		\begin{minipage}[t]{0.33\linewidth}
			\centering
			\label{fig:simu2-a}
			\includegraphics[width = 1\columnwidth,keepaspectratio]{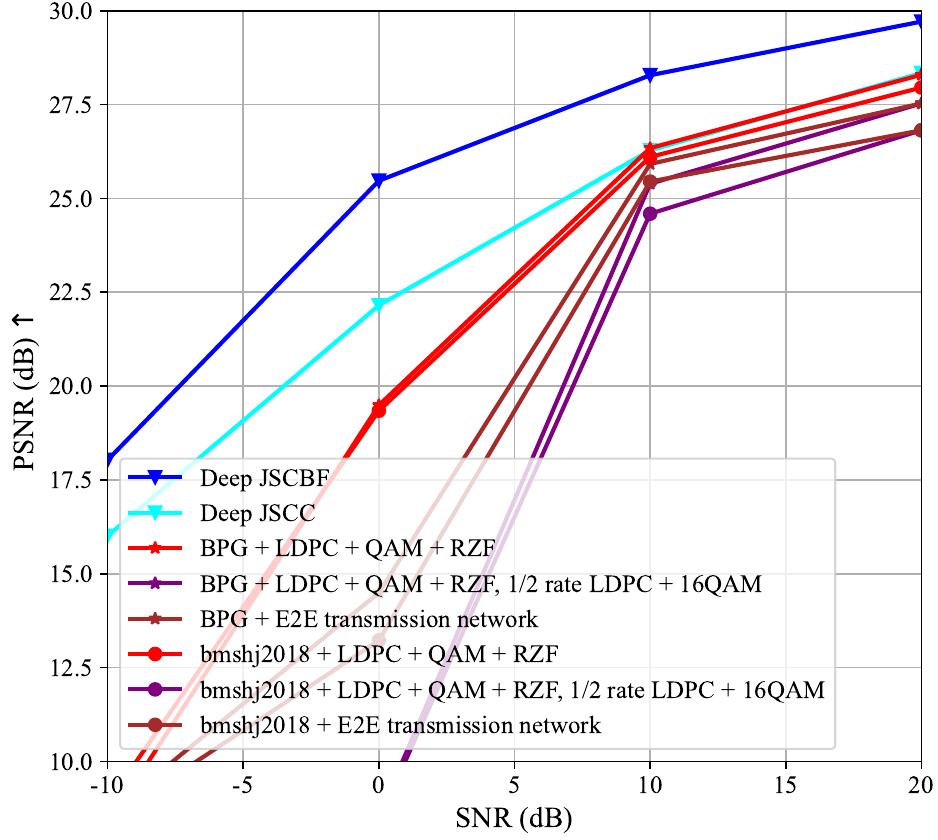}\\
		\end{minipage}%
	}%
	\subfigure[]{
		\begin{minipage}[t]{0.33\linewidth}
			\centering
			\label{fig:simu2-b}
			\includegraphics[width = 1\columnwidth,keepaspectratio]{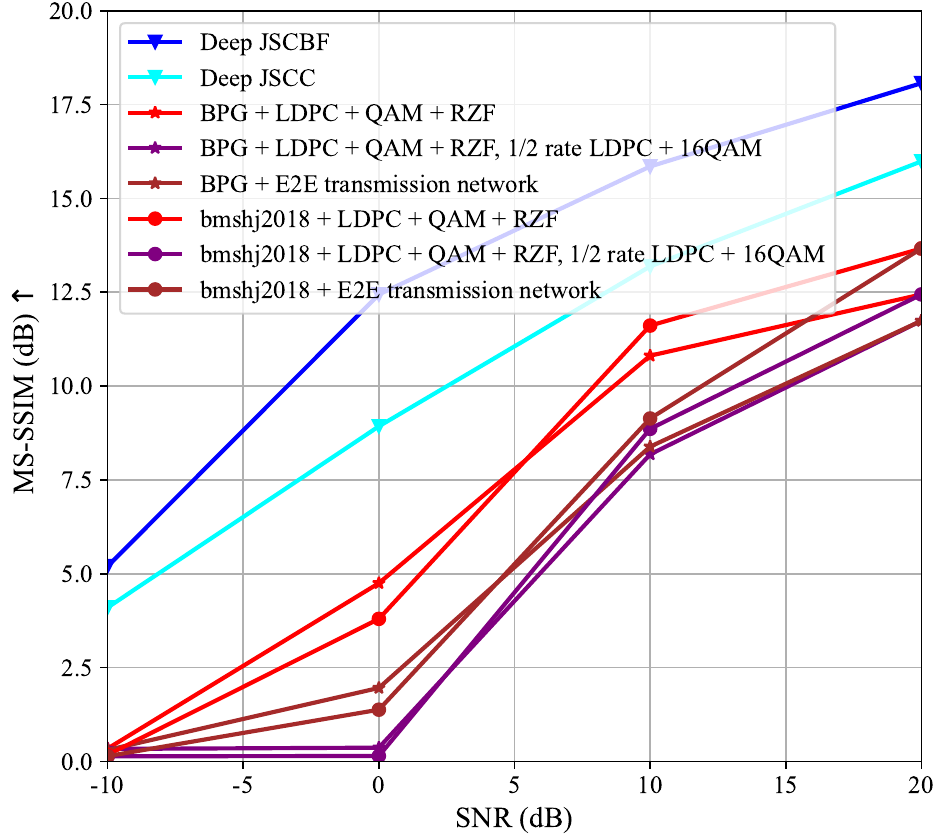}\\
		\end{minipage}%
	}%
	\subfigure[]{
		\begin{minipage}[t]{0.33\linewidth}
			\centering
			\label{fig:simu2-c}
			\includegraphics[width = 1\columnwidth,keepaspectratio]{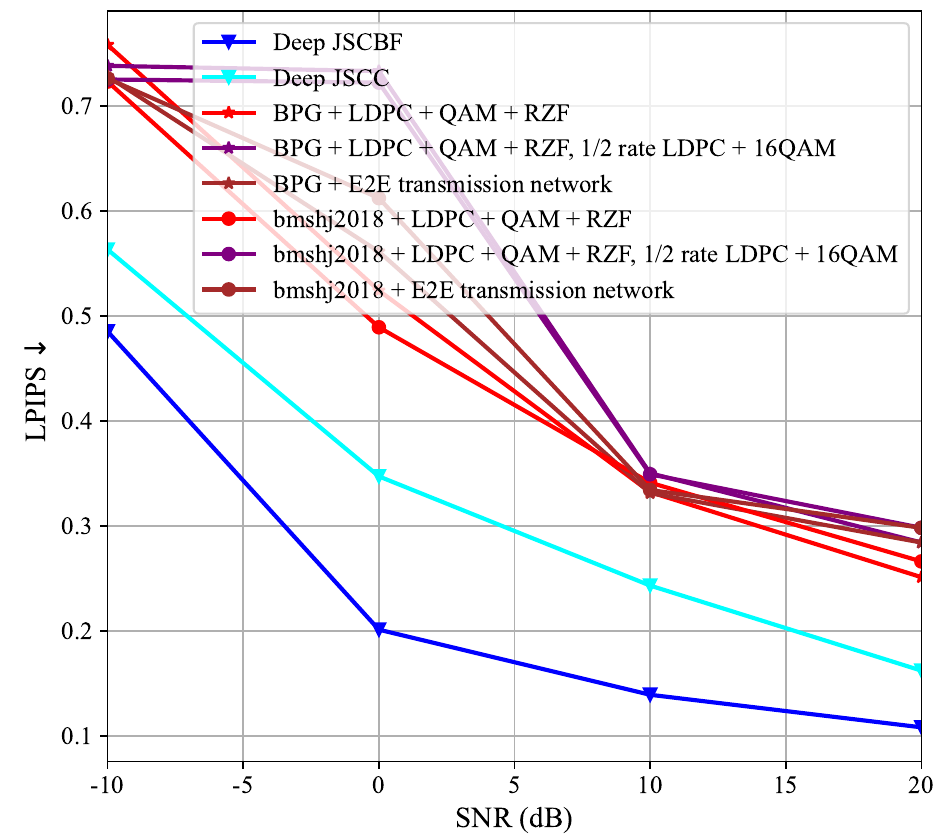}\\
		\end{minipage}%
	}%
	\centering
	\setlength{\abovecaptionskip}{-1mm}
	\captionsetup{font={footnotesize}, singlelinecheck = off, justification = raggedright,name={Fig.},labelsep=period}
\caption{Performance comparison of different solutions versus SNR at $L = 32$, given $Q = 128$: (a) PSNR performance comparison; (b) MS-SSIM performance comparison; (c) LPIPS performance comparison. }
	\label{fig:simu2} 
\vspace*{-1mm}
\end{figure*}

\subsection{Performance Comparison}\label{S6.4}

In Fig.~\ref{fig:simu1}, we compare the performance of the proposed deep JSCBF scheme with the baseline schemes in terms of the pixel-level distortion metric PSNR as well as the perceptual metrics MS-SSIM and LPIPS over different numbers of pilot OFDM symbols $L$, given $\text{SNR}=20$\,dB and the number of transmitted OFDM symbols $Q=128$. It is important to note that since both BPG and bmshj2018 are variable-length source coding schemes, they cannot achieve an arbitrarily specified compression rate. Therefore, we exhaustively search the quality factor of BPG and bmshj2018, as well as the LDPC code rate and the QAM modulation order, to obtain a number of transmitted OFDM symbols $Q$ that is closest to that of the proposed deep JSCBF scheme to ensure a fair comparison.

In Fig.~\ref{fig:simu1-a}, it is observed that the `BPG/bmshj2018+LDPC +QAM+RZF, 1/2 rate LDPC+16QAM' schemes experience significant performance degradation under inaccurate CSI estimation (i.e., $L\leq 8$), a phenomenon often referred to as the `cliff effect'. This is attributed to a large mismatch between the actual CSI and the estimated CSI, leading to a large number of errors that interfere with the normal functioning of the source decoding module. Although the `BPG/bmshj2018+E2E transmission network' schemes adopt a DL-based E2E physical-layer design, their source coding and physical-layer design remain separate and are thus still susceptible to the `cliff effect'. In contrast, the `BPG/bmshj2018+LDPC+QAM+RZF' schemes perform better because they can ensure image compression quality while keeping the bit error rate as low as possible by exhaustively searching the source coding compression rate, LDPC code rate, and QAM modulation order. The existing deep JSCC scheme improves the robustness to imperfect CSI with E2E training, and performs close to the `bmshj2018/BPG+LDPC+QAM+RZF' schemes. In comparison, the proposed deep JSCBF scheme jointly exploits the channel and image semantics to design a hybrid data and model-driven beamforming scheme, and integrates the beamforming module with the semantic extraction and decoding modules for joint E2E training, thereby achieving optimal performance. Especially under imprecise CSI estimation, the PSNR of the proposed deep JSCBF scheme outperforms the other baseline schemes by nearly 5\,dB. However, there is still a gap of about 2\,dB in PSNR compared to the idealized `BPG/bmshj2018+capacity'.

\begin{figure*}[!t]
	\vspace{-4mm}
	\centering
	\subfigure[]{
		\begin{minipage}[t]{0.33\linewidth}
			\centering
			\label{fig:simu3-a}
			\includegraphics[width = 1\columnwidth,keepaspectratio]{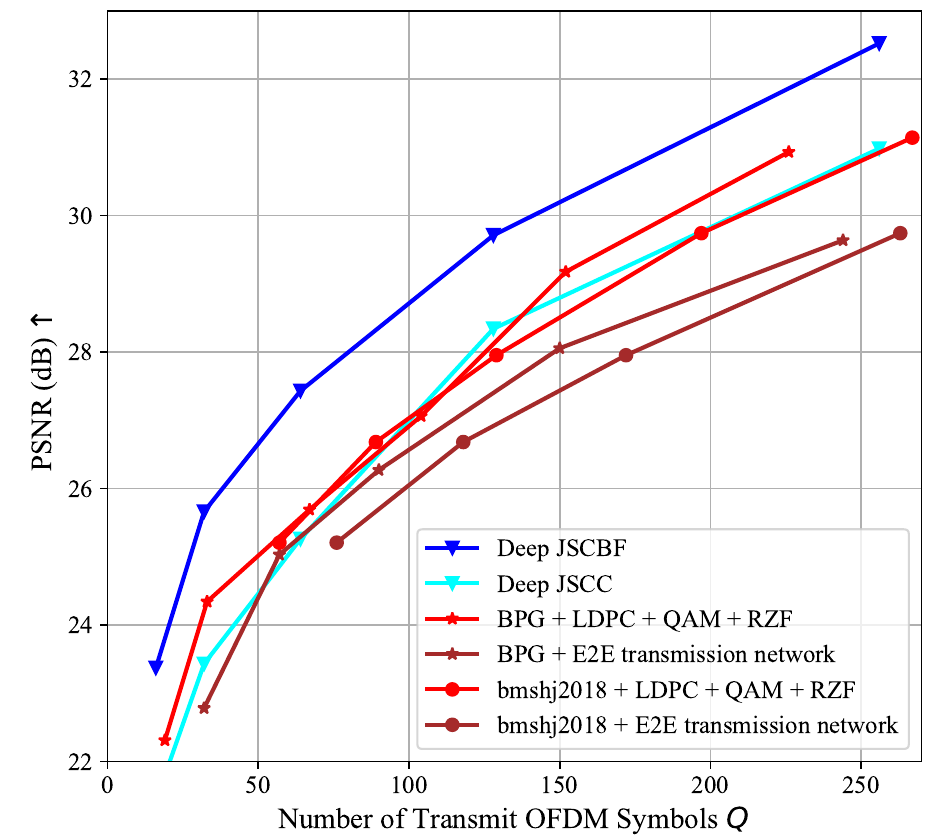}\\
		\end{minipage}%
	}%
	\subfigure[]{
		\begin{minipage}[t]{0.33\linewidth}
			\centering
			\label{fig:simu3-b}
			\includegraphics[width = 1\columnwidth,keepaspectratio]{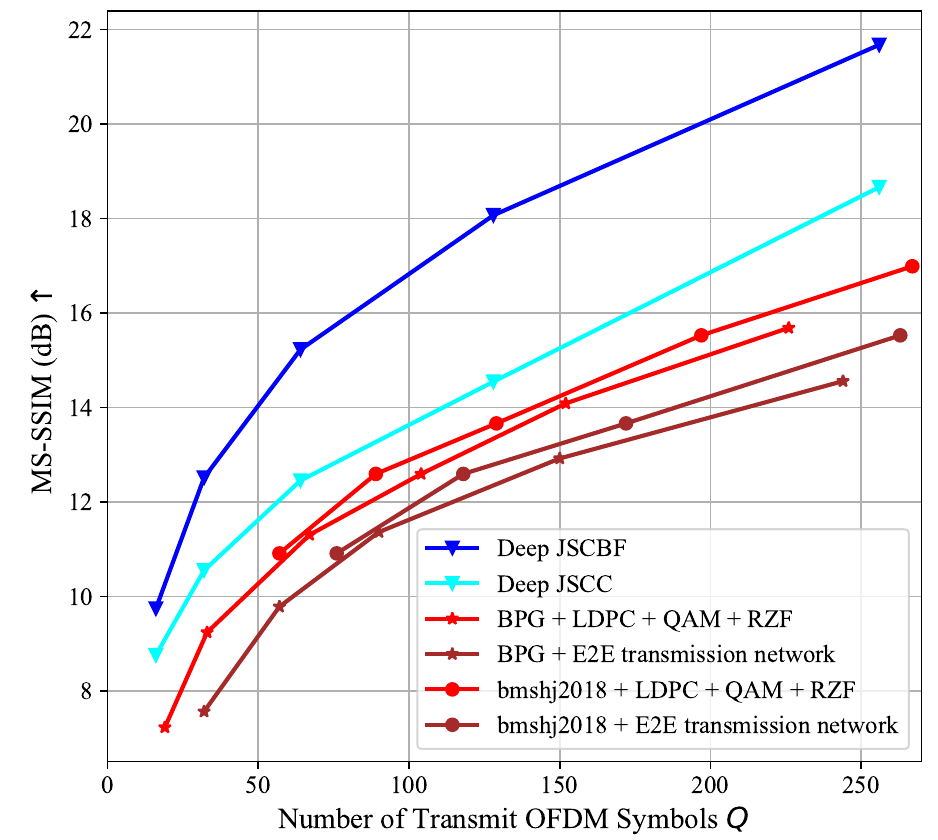}\\
		\end{minipage}%
	}%
	\subfigure[]{
		\begin{minipage}[t]{0.33\linewidth}
			\centering
			\label{fig:simu3-c}
			\includegraphics[width = 1\columnwidth,keepaspectratio]{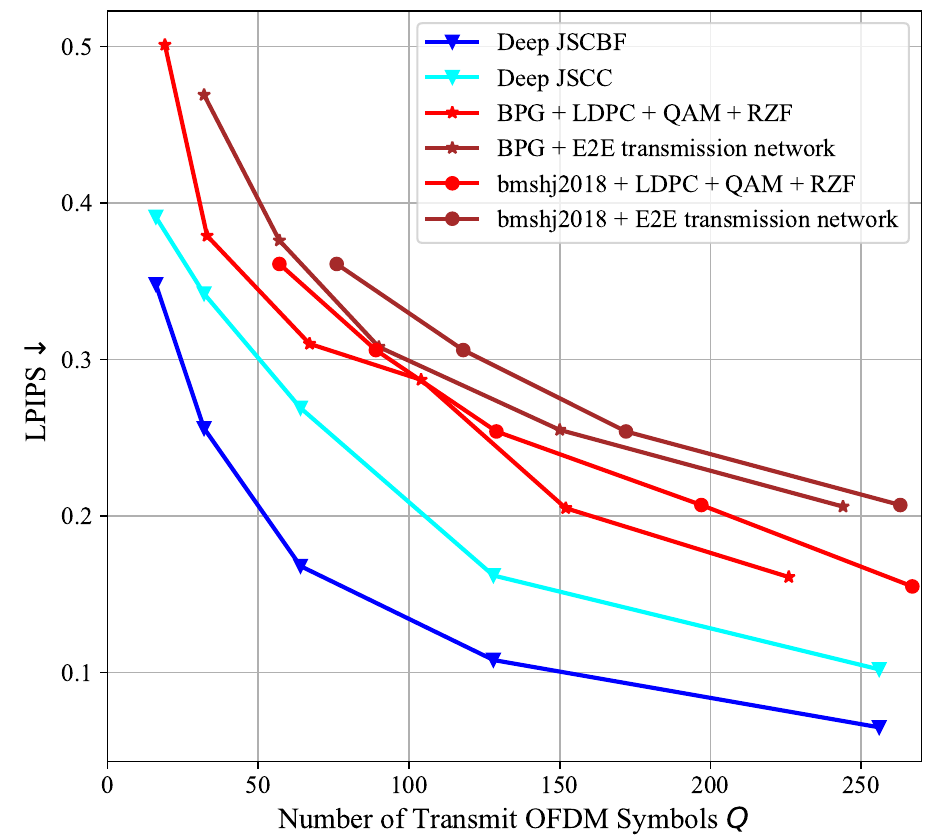}\\
		\end{minipage}%
	}%
	\centering
	\setlength{\abovecaptionskip}{-1mm}
	\captionsetup{font={footnotesize}, singlelinecheck = off, justification = raggedright,name={Fig.},labelsep=period}
\caption{Performance comparison of different solutions versus transmit OFDM symbols $Q$, given $\text{SNR}=20$\,dB and $L = 32$: (a) PSNR performance comparison; (b) MS-SSIM performance comparison; (c) LPIPS performance comparison.}
	\label{fig:simu3} 
	\vspace*{-6mm}
\end{figure*}

\begin{figure*}[!b]
	\vspace{-5mm}
	\centering
	\subfigure[]{
		\begin{minipage}[t]{0.33\linewidth}
			\centering
			\label{fig:simu4-a}
			\includegraphics[width = 1\columnwidth,keepaspectratio]{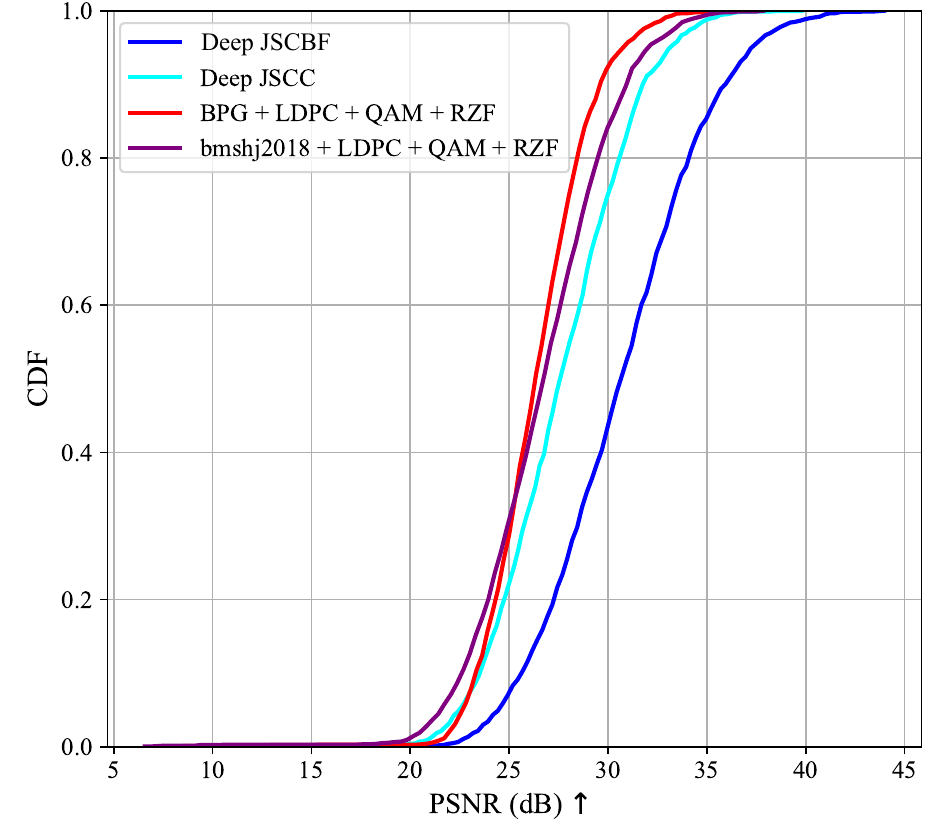}\\
		\end{minipage}%
	}%
	\subfigure[]{
		\begin{minipage}[t]{0.33\linewidth}
			\centering
			\label{fig:simu4-b}
			\includegraphics[width = 1\columnwidth,keepaspectratio]{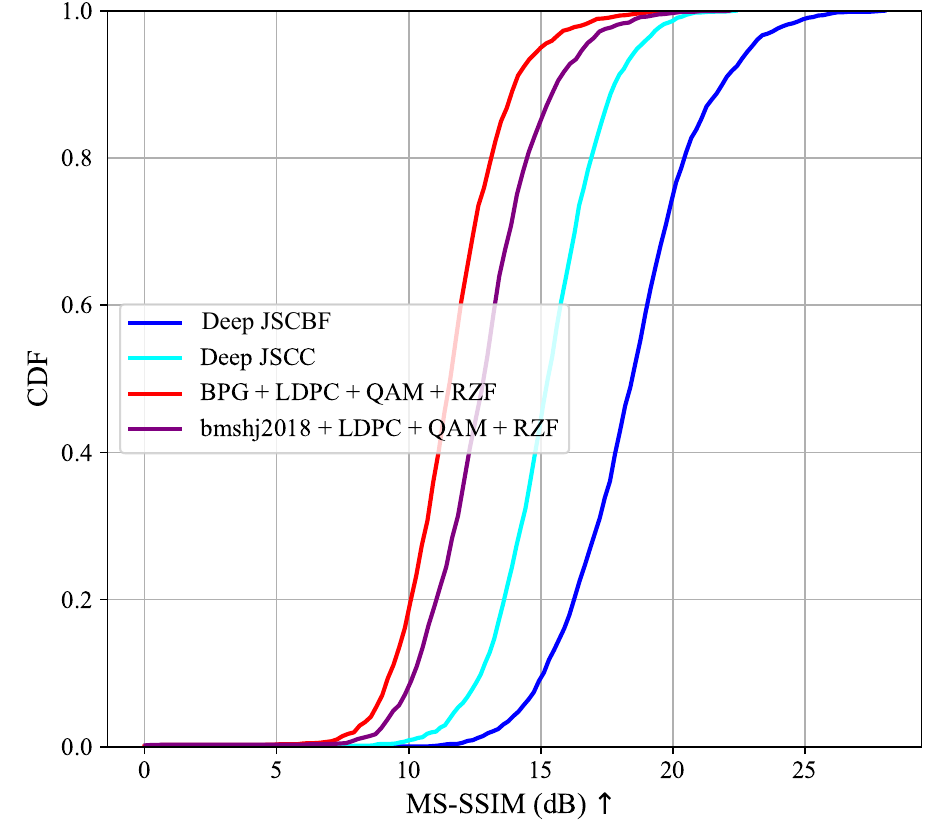}\\
		\end{minipage}%
	}%
	\subfigure[]{
		\begin{minipage}[t]{0.33\linewidth}
			\centering
			\label{fig:simu4-c}
			\includegraphics[width = 1\columnwidth,keepaspectratio]{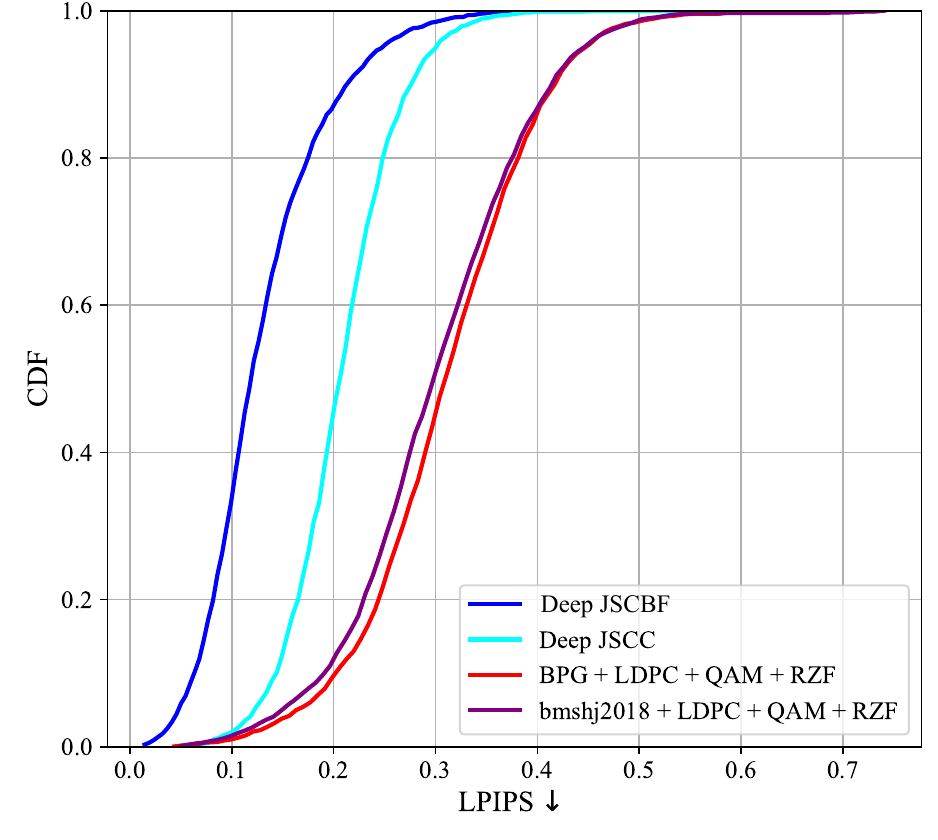}\\
		\end{minipage}%
	}%
	\centering
	\setlength{\abovecaptionskip}{-1mm}
	\captionsetup{font={footnotesize}, singlelinecheck = off, justification = raggedright,name={Fig.},labelsep=period}
\caption{The CDFs of the performance distributions for the reconstructed images achieved by different schemes, given $Q=128$, $L=16$ and $\text{SNR}=20$\,dB: (a) PSNR performance comparison; (b) MS-SSIM performance comparison; (c) LPIPS performance comparison.}
	\label{fig:simu4} 
	\vspace{-1mm}
\end{figure*}

It can be seen from Fig.~\ref{fig:simu1-b} and Fig.~\ref{fig:simu1-c} that the existing deep JSCC scheme has a significant advantage over traditional separate design approaches in perceptual metrics such as MS-SSIM and LPIPS. 
The proposed deep JSCBF scheme exhibits the best perceptual performance. In particular, the MS-SSIM of the proposed deep JSCBF scheme exceeds that of the separate design approaches by more than 5\,dB, and the LPIPS of the proposed deep JSCBF scheme is less than half that of the separate design approaches. In addition, compared to the idealized `BPG/bmshj2018 + capacity', the proposed deep JSCBF scheme exhibits better MS-SSIM performance and significantly better LPIPS performance. This is because traditional source coding optimizes images based only on pixel-level distortion, which fails to adequately extract semantic information that is more relevant to perceptual metrics. In contrast, the proposed deep JSCBF scheme, through the E2E training process, can extract semantic information most relevant to a loss function composed of pixel-level distortion loss ($\mathcal{L}_{\rm MSE}$) and perceptual losses ($\mathcal{L}_{{\rm MS}-{\rm SSIM}}$ and $\mathcal{L}_{\rm LPIPS}$), filtering out redundant information. This approach balances image pixel-level distortion performance with perceptual performance, while reducing image transmission overhead. Therefore, although the proposed deep JSCBF scheme may not perform as well as the idealized `BPG/bmshj2018 + capacity' in terms of pixel-level distortion, it has a significant advantage in perceptual metrics.

Fig.~\ref{fig:simu2} shows the performance as a function of SNR achieved by the different schemes. 
It can be seen that the traditional separate design approaches experience significant performance degradation under low SNR conditions. In contrast, the deep JSCC scheme, with its integrated optimization of source and channel coding, shows superior robustness to low SNR compared to the traditional separate designs. In particular, the proposed deep JSCBF scheme achieves the best performance.
In terms of PSNR, as shown in Fig.~\ref{fig:simu2-a}, the proposed deep JSCBF scheme demonstrates a gain of about 2\,dB over the traditional separate design approaches at high SNR levels (i.e., $\text{SNR} \geq 10$\,dB), while it outperforms the traditional separate design approaches by more than 5\,dB at low SNR levels, highlighting its robustness to challenging signal conditions. In addition, the proposed deep JSCBF scheme consistently exhibits a 2\,dB improvement in PSNR over all SNR levels compared to the deep JSCC, which is mainly attributed to the optimization of the physical layer beamforming in our approach.
In terms of perceptual metrics, as shown in Fig.~\ref{fig:simu2-b} and Fig.~\ref{fig:simu2-c}, the proposed deep JSCBF scheme exhibits significant improvements over the conventional separate design approaches. The MS-SSIM of our scheme surpasses that of the separate designs by more than 5\,dB, and its LPIPS is less than half that of the separate designs. Furthermore, the perceptual metrics of the proposed deep JSCBF scheme are consistently better than the deep JSCC at all SNR levels. These results confirm the effectiveness of the proposed deep JSCBF scheme in exploiting semantically aware hybrid data and model-driven beamforming for improved image reconstruction performance. 

\begin{figure*}[!t]
	\vspace*{-5mm}
	\centering
	\includegraphics[width = 2 \columnwidth,keepaspectratio]
	{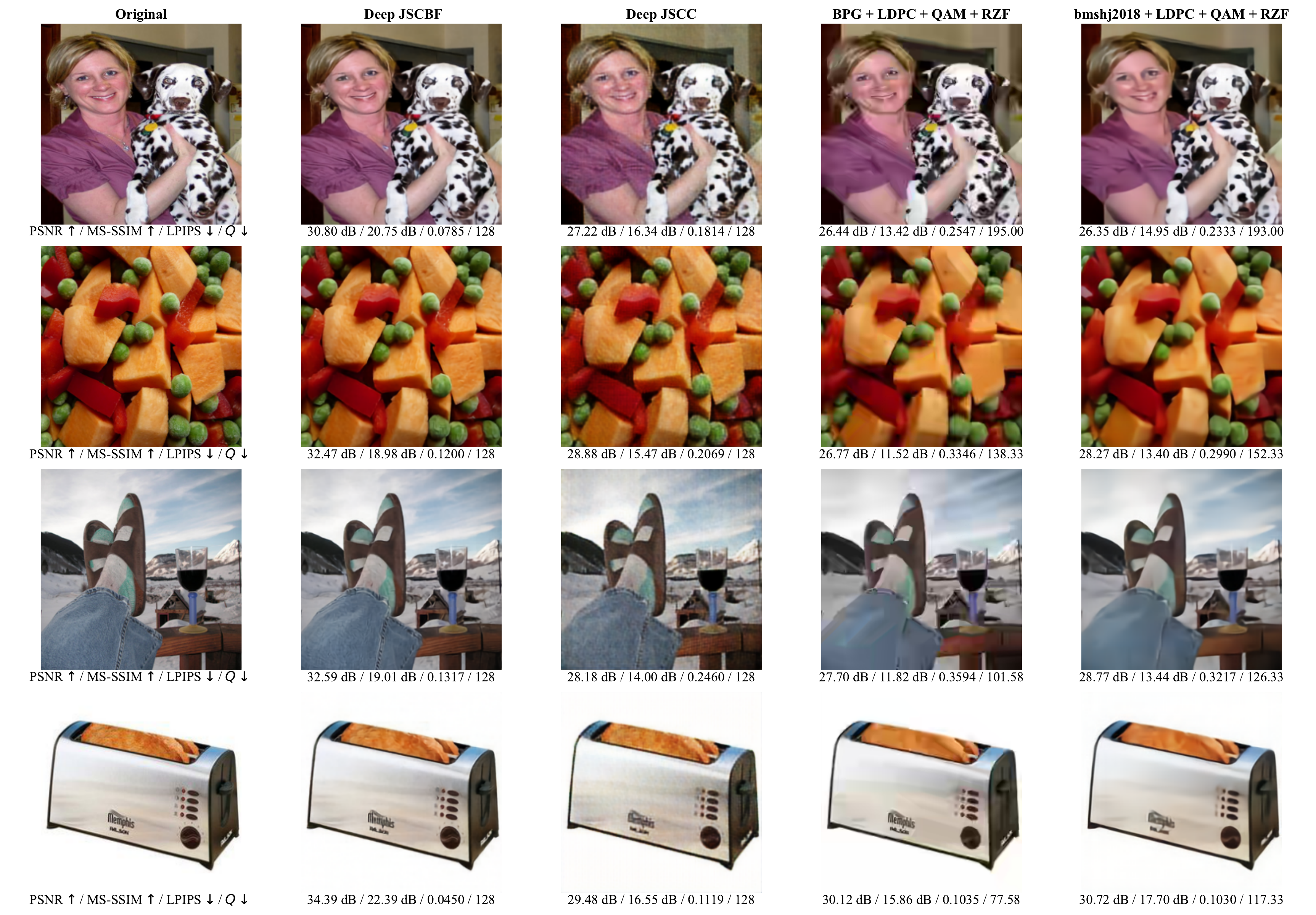}
	\captionsetup{font={footnotesize}, singlelinecheck = off, justification = raggedright,name={Fig.},labelsep=period}
	\vspace*{-3mm}
	\caption{Examples of visual comparison for different schemes.}
	\label{fig:simulation_grid_with_metrics} 
	\vspace*{-7mm}
\end{figure*}

\begin{table*}[!b]
\vspace*{-4mm}
	\centering 
	\color{black}
	\captionsetup{font={color = {black}}}
	\caption{Ablation study.}
	\vspace*{-2mm}
	\label{table2}  
	\begin{tabular}{|ll|l|l|l|l|l|}
		\hline
		\multicolumn{2}{|l|}{Number of pilot OFDM symbols $L$}                                                                                                                                                         & 4                 & 8                 & 16                & 32                & 64                \\ \hline
		\multicolumn{1}{|l|}{\multirow{3}{*}{\begin{tabular}[c]{@{}l@{}}PSNR (dB)/\\ MS-SSIM (dB)/\\ LPIPS\end{tabular}}} & Deep JSCBF                                                      & 24.10/12.07/0.254 & 27.41/14.43/1.188 & 29.05/17.03/0.125 & 29.71/18.0/0.113  & 29.80/19.32/0.106 \\ \cline{2-7} 
		\multicolumn{1}{|l|}{}                                                                                            & \begin{tabular}[c]{@{}l@{}}Deep JSCBF \\ (model-driven \\ beamforming only)\end{tabular} & 22.89/10.51/0.306  & 26.56/13.87/0.215 & 28.74/16.57/0.139 & 29.54/17.81/0.121 & 29.63/19.15/0.112 \\ \cline{2-7} 
		\multicolumn{1}{|l|}{}                                                                                            & Deep JSCC                                                                                  & 19.09/5.56/0.392  & 24.91/12.28/0.249 & 27.33/14.58/0.194 & 28.04/15.98/0.170 & 28.15/16.67/0.162 \\ \hline
	\end{tabular}
\vspace*{-1mm}
\end{table*}

Fig.~\ref{fig:simu3} shows the performance of the different schemes as a function of the number of transmitted OFDM symbols $Q$, where a smaller $Q$ implies the use of a higher compression rate to compress the images, thereby reducing the communication overhead. It is evident that for all values of $Q$, the proposed deep JSCBF scheme significantly outperforms both the separate design approaches and deep JSCC in terms of pixel-level distortion metrics and perceptual metrics. Therefore, to achieve the same level of performance, our deep JSCBF scheme imposes significantly lower communication overhead than the other schemes.

Fig.~\ref{fig:simu4} shows the cumulative distribution functions (CDFs) of PSNR, MS-SSIM, and LPIPS performance for the reconstructed images under the different schemes, given $L\! =\! 16$, $Q\! =\! 128$ and $\text{SNR}\! =\! 20$\,dB. Unlike the previous simulations, which focus only on the average power, this analysis considers the power distribution of all samples in the data set, providing a more comprehensive understanding beyond the aggregate average power. It is observed from Fig.~\ref{fig:simu4-a} that with the application of the proposed deep JSCBF scheme, approximately 80\% of the reconstructed images exceed a PSNR of 27dB, while less than 50$\%$ of the images reconstructed by the other baseline schemes achieve this performance level. 
In terms of perceptual metrics, as can be seen from Fig.~\ref{fig:simu4-b} and Fig.~\ref{fig:simu4-c}, 80\% of the images reconstructed by the proposed deep JSCBF scheme exceed an MS-SSIM of 17 dB and maintain an LPIPS loss below 0.17, while only about 20\% of the images reconstructed by the deep JSCC scheme meet this benchmark, and even fewer, about 5\%, of the images reconstructed by the other separate design approaches reach this level. These observations attest to the ability of the proposed deep JSCBF scheme to reconstruct images with a high probability of significantly superior performance compared to traditional schemes.

Finally, Fig.~\ref{fig:simulation_grid_with_metrics} presents a visual comparison of the reconstructed images obtained by the different schemes, given $L\! =\! 16$, $Q\! =\! 128$, and $\text{SNR}\! =\! 20$\,dB. It can be seen that the visual representation of the images reconstructed by the proposed deep JSCBF scheme has higher clarity and better detail representation compared to the baseline schemes. This further confirms the fact that the proposed deep JSCBF scheme has significant advantages over the other baseline approaches, both in terms of pixel-level distortion and perceptual metrics. 

\subsection{Ablation Study}\label{S4.5}

To illustrate the purpose of using a semantic-aware hybrid of data and model-driven beamforming, and to investigate the impact of estimated CSI, Table~\ref{table2} presents relevant ablation experiments. Here, `deep JSCBF (model-driven beamforming only)' refers to a simplified version of the proposed deep JSCBF scheme with the semantic-aware data-driven beamforming module removed. It is observed that the proposed deep JSCBF scheme, when using model-driven beamforming only, still exhibits performance advantages over the traditional deep JSCC. This is attributed to the use of semantic information to optimize the key parameter set $\mathcal{A}$ in beamforming, thus better addressing the performance degradation problem caused by the use of CSI estimation. The model-driven beamforming approach that integrates expert knowledge can achieve satisfactory performance when the number of OFDM pilot symbols is sufficient. However, in the case of insufficient number of pilot OFDM symbols, the model-driven beamforming approach suffers from significant performance degradation due to inaccurate CSI estimation. In contrast, the proposed deep JSCBF scheme builds on the model-driven approach by adding an additional data-driven branch to compensate for the shortcoming of the model-driven branch. This exploitation of the data-driven method enables the proposed deep JSCBF to achieve better performance under inaccurate CSI estimation.

\section{Conclusions}\label{S7}

This paper has proposed a novel deep JSCBF approach for airship-based massive MIMO near space image transmission network, where the semantic encoding/decoding and MIMO beamforming modules are collectively modeled as a unified E2E neural network, which includes image and CSI semantic extraction networks, a semantic fusion network, hybrid data and model-driven semantic-aware beamforming networks, and a semantic decoding network. Specifically, we have built the image and CSI semantic extraction networks based on the transformer architecture to extract semantics from both image source data and CSI, which are used to support the subsequent semantic fusion and beamforming.
A semantic fusion network has been developed to fuse the semantics of image source data and CSI to form complex-valued semantic features for subsequent physical layer transmission. To balance the advantages of data-driven and model-driven DL, we further designed the hybrid data and model-driven semantic-aware beamforming networks. The results of these two beamforming networks are then weighted and merged to improve the beamforming performance.
Finally, a semantic decoding network based on the Swin Transformer architecture was employed at the UE side to reconstruct images from the received signals. We have performed E2E joint training for all the modules using a loss function that combines MSE, MS-SSIM, and LPIPS.
Numerous simulation results have demonstrated that the proposed deep JSCBF scheme significantly outperforms existing separation module design schemes as well as the existing deep JSCC, especially in the case of low SNR or insufficient pilot overhead.

\bibliographystyle{IEEEtran}
	

\end{document}